\documentclass[aps,pra,twocolumn,10pt,superscriptaddress,longbibliography]{revtex4-1}
\usepackage[utf8]{inputenc}  
\usepackage[T1]{fontenc}     
\usepackage[british]{babel}  
\usepackage[scaled=0.86]{berasans}  
\usepackage[colorlinks=true,allcolors=blue]{hyperref}  
\usepackage{graphicx} 
\usepackage[babel]{microtype}  
\usepackage{amsmath,amssymb,amsthm,bm,amsfonts,mathrsfs,bbm} 
\usepackage{setspace}
\usepackage{braket}
\usepackage{blochsphere}
\usepackage{pgfplots}
\usepackage{float}
\usetikzlibrary{calc,3d,shapes, pgfplots.external, intersections}
\usepackage[bitstream-charter]{mathdesign}
\usepackage{tikz}
\usepackage{mathtools}
\usetikzlibrary{arrows.meta}
\usepackage{csquotes}
\usepackage{multirow}
\usepackage{adjustbox}

\usepackage{xspace}  
\usepackage{pgfplots}
\usepackage{verbatim}
\usepackage[bitstream-charter]{mathdesign}
\usepackage{todonotes}
\usepackage{xfrac}

\usepackage{array, tabularx, caption, boldline}
\usepackage{cellspace}
\usepackage{makecell}

\newcommand\id{\leavevmode\hbox{\small1\kern-3.3pt\normalsize1}}

\begin{document}


\title{qkdSim: An experimenter's simulation toolkit for QKD with imperfections, and its performance analysis with a demonstration of the B92 protocol using heralded photons}

\author{Rishab Chatterjee}
\affiliation{Raman Research Institute, C. V. Raman Avenue, Sadashivanagar, Bengaluru, Karnataka 560080, India}
\author{Kaushik Joarder}
\affiliation{Raman Research Institute, C. V. Raman Avenue, Sadashivanagar, Bengaluru, Karnataka 560080, India}
\author{Sourav Chatterjee}
\affiliation{Raman Research Institute, C. V. Raman Avenue, Sadashivanagar, Bengaluru, Karnataka 560080, India}
\author{Barry C. Sanders}
\affiliation{Raman Research Institute, C. V. Raman Avenue, Sadashivanagar, Bengaluru, Karnataka 560080, India}
\affiliation{Institute for Quantum Science and Technology, University of Calgary, Alberta T2N 1N4, Canada}
\author{Urbasi Sinha}
\email[]{usinha@rri.res.in}
\affiliation{Raman Research Institute, C. V. Raman Avenue, Sadashivanagar, Bengaluru, Karnataka 560080, India}

\begin{abstract}

Quantum Key Distribution (QKD) is one of the most important aspects of quantum cryptography. Using laws of quantum mechanics as the basis for security, the key distribution process is made information theoretically secure in QKD. With the advancement and commercialization of QKD, an end-to-end QKD simulation software is required that can include experimental imperfections. Software of this kind will ensure that resources are invested only after prior performance analysis, and is faithful to experimental capacities and limitations. In this work, we introduce our QKD simulation toolkit qkdSim, which is ultimately aimed at being developed into such a software package that can precisely model and analyse any generic QKD protocol. We present the design, implementation and testing of a prototype of qkdSim that can accurately simulate our own experimental demonstration of the B92 protocol. The simulation results match well with experiment; a representative key rate and QBER from experiment is $51 \pm 0.5$ Kbits/sec and $4.79\% \pm 0.01\%$ respectively, wherein the simulation yields $52.83 \pm 0.36$ Kbits/sec and $4.79\% \pm 0.01\%$ respectively.

\end{abstract}

\maketitle


\section{Introduction}

Quantum key distribution (QKD) is a promising technology that allows two distant parties, popularly referred to as Alice (sender) and Bob (receiver), to share a sequence of secret bits called the \enquote{key}\,\cite{scarani2009security}.  Unlike the state-of-the-art classical public-key cryptosystems, developed on Rivest-Shamir-Adleman (RSA) algorithm\,\cite{rivest1983cryptographic}, which depend on computational security i.e.
hardness of factoring, the security of QKD systems rely on the laws of quantum physics where eavesdropping introduces detectable errors\,\cite{scarani2009security, gisin2002quantum}. 
The secret key generated from a QKD protocol implementation can provide information theoretically secure communication when the message is encrypted with the one-time pad symmetric key algorithm\,\cite{hirota2014correct, vernam1926cipher}. \par

With QKD being commercialized\,\cite{khan2018satellite, bqn}, 
sophisticated engineering techniques are being 
developed\,\cite{cao2019sdqaas, fan2018afterpulse}. To include and evaluate these growing techniques in the realization of QKD protocols, there exists a requirement of finding a cost-efficient approach. An useful alternative to the development followed by testing of an actual experimental implementation is to design a QKD simulation toolkit, 
that can accurately model the experimentation of the existing QKD protocols and deliver the required analysis.\par 

In this article, we present the design, implementation and testing of a prototype of a QKD simulation toolkit (qkdSim) which we have developed, that can simulate an experimental demonstration of the B92 protocol, while taking into account experimental imperfections. The prototype has been designed with the vision that in future it will be ultimately developed into a complete software package that can precisely model and analyze any generic QKD protocol.\par 

Earlier theoretical research has been performed on  analysis of QKD protocols\,\cite{trinh2018design, trinh2016performance} , and to model real-world QKD systems\,\cite{mailloux2015modeling}. Early stage research activities were limited to idealistic design of QKD  protocols\,\cite{zhao2008event, niemiec2011quantum, mogos2015quantum} and they also considered only a few optical components\,\cite{buhari2012efficient, halip2014simulation, buhari2012bb84, QiaoChen, JASIM2015701}. Web-based tool-kits that simulate QKD basics primarily for educational purposes have also been developed\,\cite{webbasedqkdtool, DemoBB84, NumericalQKD}. The other development in QKD-related modelling is in the simulation of QKD networks\,\cite{QKDNetSim, 8909806}. The emphasis in these QKD network simulations is on the network structure and the dissemination of the secure key through various levels of the network and not so much on the accurate simulation of the QKD protocol towards the generation of the key\,\cite{QKDNetSim, QuantumSafe, OpenQKDNetwork, 8909806}.\par
Among the existing QKD modelling frameworks, a proper modular structure and detailed modelling architecture for the design, implementation and analysis of a full system-level model can only be found in \enquote{qkdX}\,\cite{mailloux2015modeling}. Although, qkdX has been deployed as a complete software package with a modular architecture, limited attention 
has been invested in it towards modelling the actual physical processes including photon sources, from first principles, as well as detection module. Additionally, qkdX leaves behind a large room for improvement from the implementational viewpoint in the level of imperfections considered while modelling some of its optical components.\par 
With this perspective, we have developed qkdSim that supports 
quick, easy and precise simulation of physical processes and evaluation of QKD systems, while considering realistic experimental imperfections at a greater detail. For instance, while in nearly all of the earlier works, the modelling of the input photon in the QKD protocol was considered as a sequence of events, the corresponding input of our simulation is a sequence of time stamping data that follows an ideal single photon distribution (sub-poissonian, anti-bunching, $g^{2}(\tau=0) = 0$). We have also modelled the background noise as a thermal source, in order to perform more realistic error analysis. In addition to that, our detection module includes all imperfections like dead time, quantum efficiency, timing resolution as well as afterpulsing. Thus, qkdSim aims to simulate the key generation rate as well as the quantum-bit-error-rate (QBER) for a wide range of QKD protocols while taking into account an exhaustive list of experimental imperfections. This will help in filling the gap between abstract simulations of QKD and the actual experimental performance. With a more accurate prediction of experimental performance, resources may be more confidently allocated towards real-world QKD implementations. \par

Our manuscript is organized as follows. In Section\,II, we discuss some elements related to the historical developments in QKD and thereafter we introduce the general stages as well as the evaluation methodology of a QKD protocol. In Section\,III, we discuss the software process models that have been used to develop our QKD simulator. In Section\,IV, we present a detailed description of our free-space based experimental demonstration of the B92 protocol. In Section\,V, we discuss the various modules that have been constructed for the simulation of the B92 protocol implementation. In Section\,VI, we analyse the modelling of the associated physical processes including single photon generation and time-stamping of the photon arrivals. In Section\,VII, we highlight the modelling of different optical and electrical components that were used in the experimentation. In Section\,VIII, we analyze and evaluate the results simulated with qkdSim against those obtained from the actual experimental implementation using the same setup. Lastly, in Section\,IX, we provide the concluding remarks and discuss the future research efforts that can be made in this direction. The detailed analytical expressions and methodologies 
used at various stages in the toolkit, have been provided as appendices. 

\section{Background, general approach \& performance analysis of QKD systems}
\label{sec:background}

In this section, we first provide a brief summary of the historical advancements in QKD. Thereafter in the second part, we outline the main steps required to distill a secure key in any QKD protocol. Finally in the third part, we define the criteria to evaluate its security.

The first QKD protocol (BB84) was proposed in 1984 by Bennett and Brassard\,\cite{bennett1984ieee} 
and then experimentally demonstrated in 1989 over a 30 cm free-space optical channel\,\cite{bennett1992experimental}. The information theoretic security of the BB84 protocol has been proven\,\cite{gisin2002quantum, lo1999unconditional, scarani2009security, gottesman2004security}. Unlike the use of four non-orthogonal states in BB84, 
QKD was achieved in 1992 using two 
non-orthogonal states\,\cite{B92}. This was called the B92 protocol and it was experimentally realized with weak coherent pulses (WCPs) in 1998\,\cite{PhysRevA.57.2379, buttler1998practical}. Over the years, few protocols such as E91\,\cite{PhysRevLett.67.661} and BBM92\,\cite{PhysRevLett.68.557}, that perform QKD using quantum entanglement instead of the non-commutativity of quantum operators as their resource, have also been proposed and demonstrated\,\cite{erven2008entangled, ling2008e91}. In 1998, QKD was first shown to be secure  with imperfect devices, more specifically by the proposal of a self-checking photon source\,\cite{mayers1998quantum}. This work initiated developments in the topic of device-independent quantum key distribution (DIQKD) which not only proved that QKD can be fully secure with minimal fundamental assumptions and untrusted devices\,\cite{PhysRevLett.113.140501, Miller:2016:RPS:2997039.2885493, PhysRevLett.98.230501, Pironio_2009}, but also experimentally realized it\,\cite{diqkd_exp}. \par 

An important model in a QKD demonstration is the type of channel over which the key distribution is performed. Two common choices are the free-space and fibre-based ones.\par
The fibre-based 
QKD setup in 1998 was able to communicate up to 100\,km\,
\cite{buttler1998practical} following the first experimental implementation of QKD in 1992\,\cite{bennett1992experimental} and in 2003 using a free-space optical link, QKD could be demonstrated only up to 23.3\,km\,\cite{hatcher2003}. Since then, substantial progress in research  
has led to the rapid development of optical quantum technologies and over the last few years, many experiments including the demonstration of the feasibility of ground-to-satellite, satellite-to-ground and satellite-to-satellite QKD \, \cite{liao2017satellite, frohlich2017long, liu2010decoy, wang2018practical, liao2017satellite, nordholt2002practical, rarity2002ground, gunthner2017quantum, yin2017satellite, wang2013direct, liao2017long} have been reported as well as commercial QKD devices\, \cite{idquantique, magiq, qasky, quintessence} are 
available. Besides the implementation of long haul QKD\,\cite{bedington2017progress}, chip-based integrated QKD systems supporting miniaturization have been designed to enable large-scale deployment of QKD into future telecommunication networks\,\cite{sibson2017chip}.\par


The process of generating a secure key in a QKD protocol can be segregated into (i)\,authentication, (ii)\,transmission using single photons or WCPs, (iii) \,sifting, (iv)\,error correction and (v)\,privacy amplification\,\cite{nordholt2002practical, rarity2002ground}. 
More specifically, in order to obtain the secure key, the randomly generated raw key is first communicated over the quantum channel, followed by information exchange over the classical channel, that leads to the generation of the sifted key. 
Thereafter the steps (iv) and (v) are implemented. Error correction rectifies the erroneously received information bits and estimates the error rate, while finally privacy amplification extracts a shorter and even more secure final key.\par

Performance
of QKD systems are evaluated
with the QBER and the rate of secure communication\,\cite{buttler1998practical}. A lower QBER indicates higher security, while
a high secure 
communication rate implies that the transmission link has a good performance. Any information leakage to an eavesdropper about the generated key leads to an increase in the QBER. Therefore,
obtaining a high QBER value reduces the rate of secure communication during error correction stage of the QKD protocol. 
If the QBER remains below a certain threshold, then the two parties (sender and receiver) can still distill a secure key string by means of error correction and privacy amplification\,\cite{PhysRevA.72.012332}. In other words, if the QBER of the sifted key is above the information theoretically computed threshold for a given QKD protocol then the key is no longer secure. In that case, any privacy amplification technique becomes ineffective. Thus it is imperative for a QKD protocol to ensure a proper upper bound for the QBER if the privacy amplification techniques are to be employed to eliminate any knowledge gained by the eavesdropper.

                            	

\section{Software process of the QKD simulator}
\label{sec:framework}

In this section, we discuss the software development procedures on which our qkdSim has been designed. As a part of this discussion, we also highlight the salient features of those software processes and how they get associated to our final objective.\par 

A software process generally refers to the set of activities that have been used to build a software product\,\cite{sommerville2011software}. In software engineering, the simplified representation of a software process is known as a software process model or process paradigm\,\cite{sommerville2011software,pressman2005software}. Although there can be different processes and process models, each of them must satisfy four activities that are fundamental to software engineering\,\cite{sommerville2011software}. More specifically, these four activities are that (i) the software specification must be well defined, (ii) the software design and implementation must perfectly suit the requirements, (iii) the implemented software must be validated, and finally (iv) the software must posses the provision to be easily evolved as per user needs.

\subsection{Overview of the process model}

In this part, we explain the software process model that has been used to construct qkdSim. Our QKD simulation toolkit has been built using a hybrid process model\,\cite{munassar2010hybrid}.
From a top-down perspective, our version of the hybrid architectural model consists of two parts: the \enquote{Waterfall} development model which supports linear and sequential design techniques; and the \enquote{Agile} development model which allows iterative and incremental design procedures\,\cite{pressman2005software}. In this sense, it can be categorized as a kind of "Agifall" process model\,\cite{agifall}. Agifall merges the best of both worlds, by injecting Agile techniques into loose Waterfall design procedures. \par

As discussed in Section\,\ref{sec:background}, any QKD protocol grossly is a five-step sequential process, which experimentally involves propagating the signal generated at the source stage through the preparation, transmission, detection and lastly post-processing stage. Therefore, in qkdSim, the outer structure, containing the gross five-step QKD design, has been developed using the Waterfall process model, which promotes a clear \emph{flow down} logic scheme. 
However, keeping in mind the continuous and rapid evolution of technological advances, precision of handling imperfections and experimental non-idealities; the inner software development architecture for each of the five experimental stages from the modelling of components and physical processes to data processing methods have been developed using the Agile process model, which supports development at a \emph{sprinter's} pace. 
In a nutshell, by using Agifall process model we have ensured that the design pattern remains robust and modular at every step of software development to ease future customization challenges.\par


The key components from the Agifall process model, which support the \enquote{best of both worlds} logic, have been highlighted as follows\,\cite{pressman2005software}. These components have been primarily considered for the design of the inner and outer structure of qkdSim.
\begin{itemize}
\item Modelling, usability and organization can be coherently applied on the outer structure of all QKD protocols in general, so the choice of Waterfall process model is suitable.
\item Incremental updates in small update cycles allow rapid improvement of basic experimental techniques and components, so the choice of Agile process model is appropriate for designing the inner parts of the implementation stage.
\end{itemize}

\subsection{Salient features of the process model}

\begin{figure}[!ht]
    \centering
    \includegraphics[width=\linewidth]{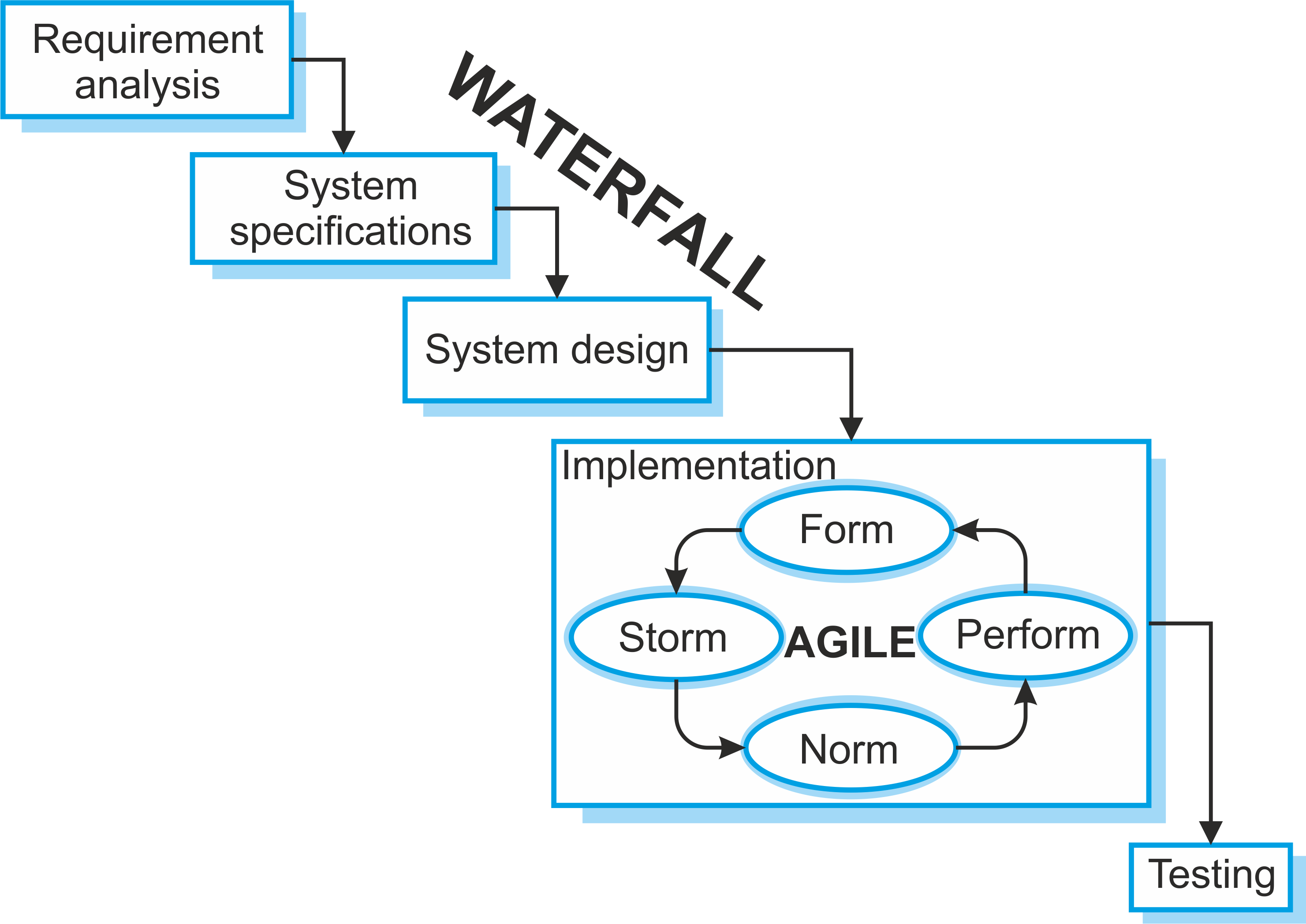}
    \caption{A schematic of the hybrid process model used to build qkdSim. The outline of the simulator has been developed on the Waterfall model, while its implementation-based  intricacies have been modelled using the Agile design procedures.}
    \label{fig:model}
\end{figure}

In this part, we first present the salient stages of our version of the Agifall process model in Figure\,\ref{fig:model}\,\cite{pressman2005software}.


\noindent Thereafter, in the following subsections, we discuss in details the five major steps of the top Waterfall structure; including the requirement analysis performed for developing the toolkit, followed by the specifications of the system, the design process and finally the implementation  and testing stage. 

\subsubsection{Requirement analysis}

The first stage of our architectural model is \enquote{requirement analysis} where the goals and constraints of the software are discussed by consultation with the users. These user requirements then serve as a part of the software specifications. The probable users for the toolkit are identified
as experimental physicists and engineers working in the domain of quantum communication and cryptography. The users are required to provide necessary inputs for the toolkit to perform. The requirements of the user that  the toolkit will be able to fulfil are analysed with the help of an user story. The principle user story for the toolkit is
\begin{quote}
\emph{I, as a QKD experimentalist, want to simulate an experimental implementation of a QKD protocol and estimate the key rate, QBER and assess the security of the implementation.}   
\end{quote}
The toolkit is also required to provide flexibility to the users to simulate any experimental setup and be able to vary the choice of various components in the setup. The toolkit is built keeping in mind that the user requirements may vary with time and hence the toolkit should be extendable to accommodate modification of the various modules that form the system. An important aspect of developing the toolkit is the consideration of the security of the protocol being simulated. The simulated results corresponds to a secure implementation of the protocol where the security parameters are clearly stated. Following the stage of requirement analysis, we move on to the stage where we identify the specific inputs and outputs of the simulation toolkit such that the user requirements are satisfied.

\subsubsection{System specifications}

\begin{figure}[!ht]
    \centering
    \includegraphics[width=\linewidth]{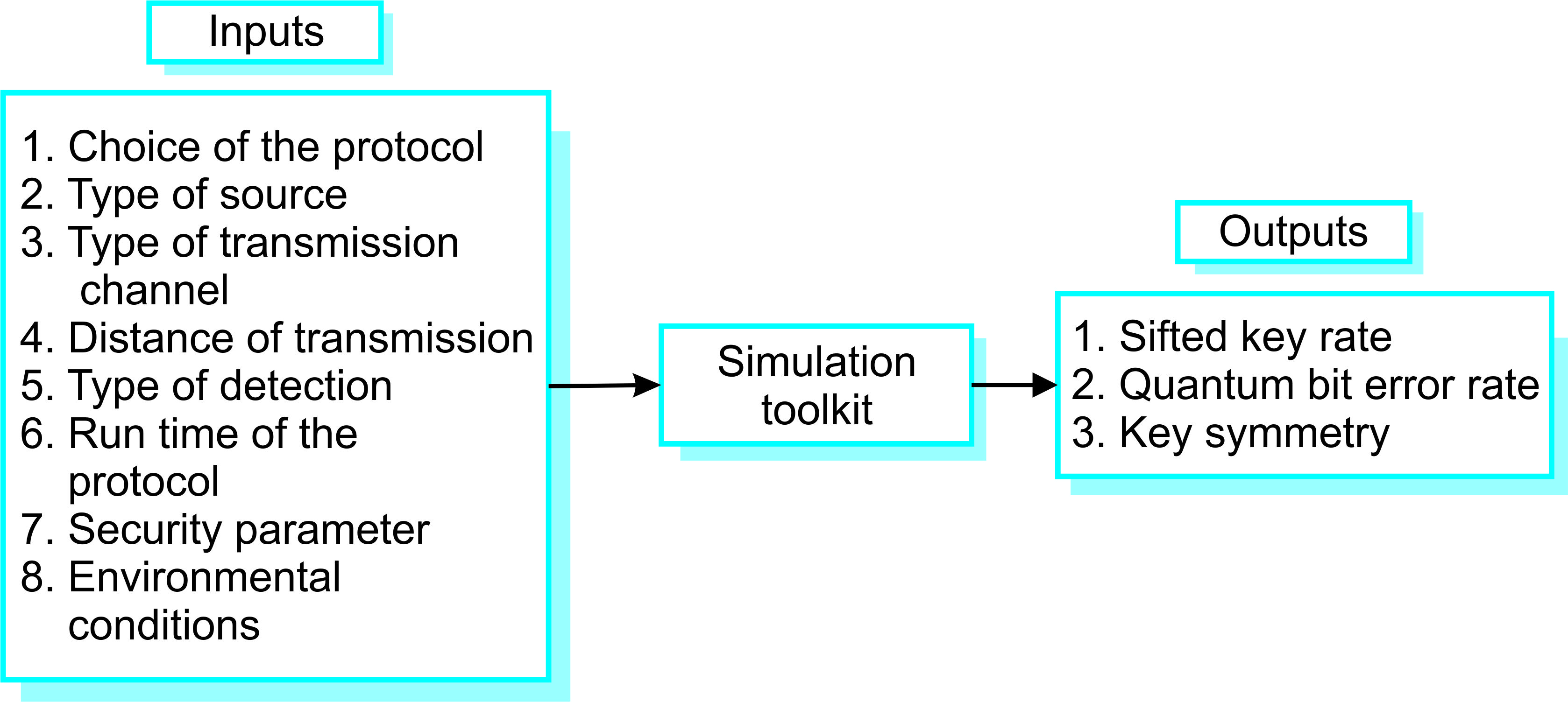}
    \caption{System specifications of the simulation toolkit}
    \label{fig:system}
\end{figure}

The second stage called \enquote{system specifications} is used to identify the inputs and outputs from the user requirements. Figure\,\ref{fig:system} represents the inputs and outputs of the simulation toolkit. At this stage, we have formed the overview of the system and identified the functions that the system is enabled to perform without delving into the question: \enquote{how}. The inputs take into account the integral components of implementation of a QKD protocol and various choices that the experimentalists posses.
The identification of the general inputs and outputs to the toolkit leads us to the design stage where we develop the architecture of the toolkit.

\subsubsection{System design}

At the third or the \enquote{system design} stage, the user requirements along with the list of inputs and outputs have to be associated with the hardware and software units to establish an overall system architecture of the toolkit.
The independent layers that will form our desired system must contain the level of abstraction and flexibility that we want to provide to the user. An optimal architecture enables us to develop the system in an iterative approach and enhance the system by considering further real-world imperfections. 
\begin{figure}[!ht]
    \centering
    \includegraphics[width=\linewidth]{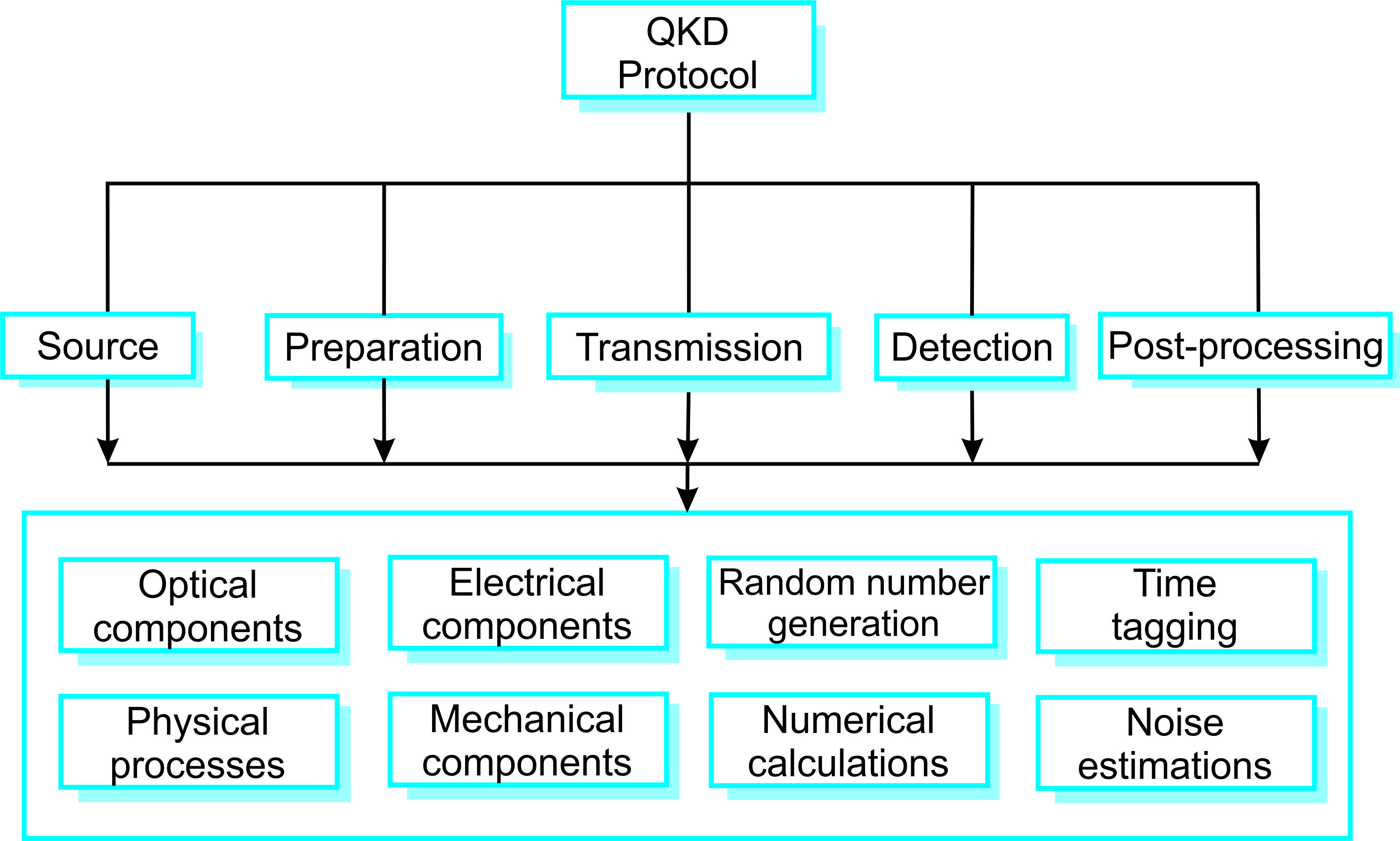}
    \caption{Architecture of the simulation toolkit.}
    \label{fig:architecture}
\end{figure}

In our design presented in Figure\,\ref{fig:architecture}, the independence of each layer of the architecture exists with respect to their development and modelling. However, there is a hierarchical dependence among the layers in forming the components of the experimental implementation to be simulated. Each layer of the architecture can be explained as follows.
\begin{itemize}
\item The QKD protocol chosen by a user forms the top layer of the system.
The choices regarding the other inputs to the toolkit will be dependent on the choice of the protocol. As an example, for an entanglement based QKD protocol, the choice of source is restricted to entangled photon sources. Thus, depending on the choice of the protocol, the respective modules from the second layer will be chosen.

\item The second layer is formed by the different modules that will be used to simulate the corresponding aspect of the experiment. The modules can be further sub-divided into different types to accommodate for the various possible requirements. For example, the source module can accommodate different types of sources such heralded single photon, weak coherent pulse source, entangled photon source, etc. Thus, the user will be able to choose the specific type corresponding to each of the modules as required for the simulation. 

\item The bottom layer is formed by the sub-modules that are modelled and are used to form the structures of the modules at the upper layer. Each of the modules will be functional by flow of logic through these sub-modules. The sub-modules contain various modelled components and processes that can be chosen by the user as required for the modules.
\end{itemize}
The different layers developed in the architecture form the basis for the next stage, which is the implementation, in an iterative manner following the Agile model mentioned previously. 

\subsubsection{Implementation}
In this work, an experimental demonstration of the B92 protocol has been simulated by using the architecture described in the previous subsection. This is a prototype of the toolkit, and the modelled physical components and processes are limited to the current aim of simulating the specific implementation. The accuracy of the prototype is limited by the various assumptions that have been considered as well as the set of imperfections that have been taken into account for modelling the physical components. 

\begin{center}
\begin{table}[!ht]
\centering
\begin{tabular}{|p{2.5cm}|p{2.5cm}|p{2.5cm}|}
\hline 
beam splitter & Polarizing beam splitter & Phase retarder\tabularnewline 
\hline 
Single mode fibre & Single photon detector & Time Correlated single photon counting module (TCSPCM)\tabularnewline
\hline 
ppKTP crystal & SMA Cable & Bandpass filter\tabularnewline
\hline 
In-lab free-space channel & Lens & Coupler \tabularnewline
\hline 
\end{tabular}
\caption{\label{tab:components}List of modelled physical components}
\end{table}
\par\end{center}

\begin{center}
\begin{table}[!ht]
\centering
\begin{tabular}{|p{2.5cm}|p{2.5cm}|p{2.5cm}|}
\hline 
Type-II SPDC process & Focussing and collimation of Gaussian beams & Fibre coupling\tabularnewline 
\hline 
Single photon detection & Time stamping & Background photon detection \tabularnewline
\hline 
\end{tabular}
\caption{\label{tab:process}List of modelled physical processes}
\end{table}
\par\end{center}

Tables\,\ref{tab:components} and\,\ref{tab:process} categorize the different physical components and processes that have been modelled and tested by considering realistic imperfections. The design methodology employed for the modelling of these two categories of elements is the Agile development procedure. Such a choice allows the required flexibility for quickly incorporating the future technological advancements, inclusion of further non-idealness and improving the considered precision levels.

The simulation toolkit currently has been implemented in Python and interacts via a command user interface. Now, we move on to the analysis and testing of the implemented prototype of the simulation toolkit.

\subsubsection{Testing}
At this stage, we test the various modules and sub-modules, that have been constructed as a part of the prototype, simulates the experimental demonstration of the B92 protocol. Each sub-module was tested to verify whether expected outcomes were obtained. The results from sub-modules corresponding to the physical components were compared with the data sets and characterization sheets of the respective components. The sub-modules of the physical processes are tested by matching the results with experimental observations corresponding to the same processes.The outputs from each of the modules were compared with the corresponding sections of the actual experimental setup. The overall verification and testing of the prototype has been done by comparing the simulated outputs with the experimental results obtained from the free-space demonstration of the protocol. In the upcoming section, that is Section\,\ref{sec:experiment} we provide a detailed discussion on the procedure of our experimental implementation, while the results obtained from the same are presented in Section\,\ref{sec:results}.

\section{Experimental demonstration of B92 protocol in free-space}
\label{sec:experiment}

In this section, we describe in details our free-space based experimental realization of the B92 protocol. In the first part, we provide a brief overview of the B92 protocol and a quick history of its various experimental implementations. In the second part, we discuss in details the experimental setup which we used to implement the B92 protocol. Lastly, we highlight the general procedure and the associated novel techniques, that we have developed and used to analyse our experimental data i.e. estimate the key rate, QBER and key symmetry for our experimental demonstration.

\subsection{General procedure of B92}


In a standard B92 protocol using polarization encoding, Alice sends Bob a stream of single photons, where the polarization state of each of the photons is randomly selected between any two non-orthogonal polarization bases, say, $\left|a\right\rangle$ and $\left|b\right\rangle$, where $\left\langle a|b\right\rangle \neq 0$. These two polarization states are encoded with binary 0 and 1, respectively. When these photons reach Bob, he randomly and independently selects between two projection operators $(I-\left|a\right\rangle\left\langle a\right|)$ or $(I-\left|b\right\rangle\left\langle b\right|)$ for each photon, and projects on it. Note that the operator $(I-\left|a\right\rangle\left\langle a\right|)$ always gives a null result when it operates on $\left|a\right\rangle$. Similarly, $(I-\left|b\right\rangle\left\langle b\right|)$ always gives a null result for $\left|b\right\rangle$. \\
\begin{figure*}
    \centering
    \includegraphics[width=\textwidth]{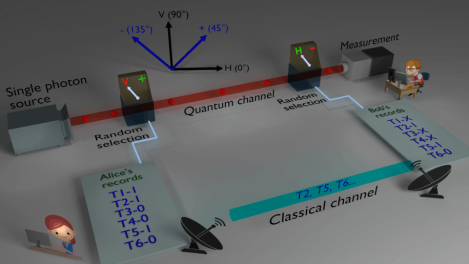}
    \caption{Schematic of the B92 protocol based on polarization encoding.}
    \label{fig:SchemeB92}
\end{figure*}
A schematic of the B92 protocol based on polarization encoding is presented in Figure\,\ref{fig:SchemeB92}. Here, Alice randomly selects the polarization state for each photon coming from a single photon source to be either vertical (V) or diagonal (+) and sends it to Bob. She also assigns bit values to all photons based on their polarization and records them sequentially (T1-1, T2-1, T3-0, etc.).  Bob randomly selects his measurement basis to be horizontal (H) or anti-diagonal (-). Bob only considers those events where his measurements give positive (or non-null) results (T2, T5, T6, etc.) and announces only the occurrence (or timing) of these positive events in a public communication channel (that may be prone to eavesdropping), once all the photons have been received. Bob never shares the choice of measurement operations for these positive events. Based on the announcement, Alice only keeps those bits that generate positive events in Bob's setup. Thus Alice and Bob generate and share an identical, secure key.\par

There have been many experimental implementations of the B92 protocol. All these experiments can be broadly categorized based on three classification parameters i.e. types of encoding, transmission medium, and type of source of photons. Though the original B92 protocol was based on phase encoding and many later experiments \cite{Bourennane:99,doi:10.1080/09500340008244058} followed similar logic, a number of experiments have also been performed based on polarization encoding \cite{PhysRevLett.81.3283,879387,4542782,inproceedings}. In terms of the transmission medium, there are experiments in free-space transmission channel \cite{PhysRevLett.84.5652,Bienfang:04,Meyers:05} as well as fibre-based channel \cite{1308613,Tang:06,4542782}. In terms of type of source, interestingly, most of the experiments till date use weak coherent pulses (WCP) as single photons \cite{Garcia-Martinez:13,PhysRevA.91.042320,doi:10.1080/09500340008244058,Tang:06}, and only a handful of experiments have considered heralded single photons generated from spontaneous parametric down-conversion (SPDC) \cite{Meyers:05,inproceedings}.\par

In our implementation, we have used heralded single photons generated using the SPDC process. Our choice of source is connected to the security aspects of the protocol and enhances the same. This is explained in detail in the paragraph on post-processing below. The photons have been encoded in polarization degree of freedom, and transmitted in a free-space channel inside a lab environment. 

\subsection{Our experimental implementation of the B92 protocol}
\label{subsec:exp}

In this part, we will provide a detailed description of the key resources and the various stages of our experimental implementation. In the process of analyzing the different stages of our setup and its associated components, we also identify the different sources of noise and imperfection, that can potentially affect our measurements, and highlight how our resources help to mitigate them.\par

We use spontaneous parametric down conversion (SPDC) as a source of heralded single photons. While there are a very small number of SPDC based implementations of B92 protocol in literature, our implementation is significantly different from the existing ones as discussed below. \par

\begin{figure*}[!ht]
    \centering
    \includegraphics[width=0.8\textwidth]{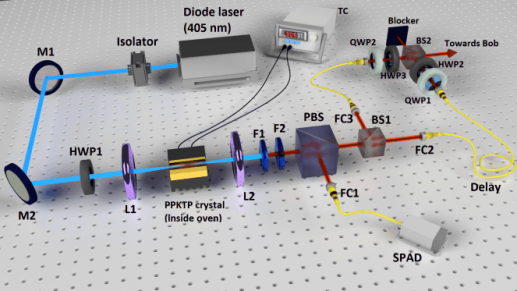}
    \caption{Schematic of the heralded single photon source and Alice's module. \textbf{M1, M2}: dielectric mirrors; \textbf{HWP1}: half-waveplate for pump light; \textbf{L1}: focusing lens; \textbf{L2}: collimating lens; \textbf{F1}: long pass filter; \textbf{F2}: band pass filter; \textbf{PBS}: polarizing beam splitter; \textbf{FC1, FC2, FC3}: fibre couplers; \textbf{BS1, BS2}: 50-50 non polarizing beam splitter; \textbf{HWP2, HWP3}: half-waveplates; \textbf{QWP1, QWP2}: quarter waveplates; \textbf{SPAD}: single photon avalanche detector; \textbf{TC}: temperature controller.  }
    \label{fig:A}
\end{figure*}

The schematic in Figure\,\ref{fig:A} has details on the components in the source. A blue diode laser of wavelength 405 nm (Cobolt 08-NLD) pumps a PPKTP crystal continuously with 30 mW power. The polarization of the pump beam is kept horizontal (H), by using a half-waveplate (HWP1). The PPKTP crystal is placed inside an oven (RAICOL), which is connected to a temperature controller. We find that at $40^{o}$ C, photon pair generation is optimal for collinear, degenerate, type-II SPDC process in the crystal,  wherein a horizontally polarized pump photon of wavelength 405 nm down-converts to a horizontally polarized signal and a vertically polarized idler photon, both with peak wavelength at 810 nm. In order to maximize photon pair generation, two lenses of focal lengths 100 mm (L1) and 50 mm (L2) respectively are used; L1 focuses the pump beam at the centre of the crystal and L2 is used to collimate the beam. Due to the collinear configuration, both photons in each pair traverse the same path as followed by the residual pump beam. A long pass filter F1, placed after the crystal, blocks most of the pump beam and allows only the red photons to go through. A band-pass filter F2, allows only photons with wavelength close to 810 nm to go through. A polarizing beam splitter (PBS) separates the two photons in each pair, where ‘H’ polarized photon goes to the transmitted arm and ‘V’ polarized photon goes to the reflected arm of the PBS. A coupler, FC1, placed in the reflected arm couples ‘V’ polarized photon to a single mode fibre that is connected to a single photon avalanche detector (SPAD; COUNT-T-100). So, detection of a ‘V’ polarized photon heralds the ‘H’ polarized photon of the same pair. This is how heralded single photons are generated.\par
\textit{Alice-substation:} The components after the PBS in Figure\,\ref{fig:A} constitute the Alice-substation of the QKD implementation. Once each heralded single photon is generated, Alice randomly selects between two non-orthogonal polarization states, vertical (V) and diagonal (D;$+45^{o}$ w.r.t. horizontal), by passing the photon through a 50:50 non-polarizing beam splitter (BS1), whose transmitted arm has a HWP (HWP2) that rotates the input H polarization to D, and reflected arm has a HWP (HWP3) that rotates the input H polarization to V. Quarter-waveplates (QWP2 and QWP3) are also placed along with HWPs, in order to minimize ellipticity in the polarization. The use of such a 50:50 BS for random selection gives us quantum randomness, that is necessary for the security aspects of the protocol. However, this comes with the caveat that Alice herself does not have knowledge about the polarization state of the photon that she send to Bob.\par

In order to resolve this, we come up with a novel solution wherein Alice makes use of two known properties of SPDC, namely, heralding process and the probabilistic nature of pair generation, where photon pairs are generated randomly in time. If Alice now randomly selects a subset from generated photon pairs and applies a fixed time delay to them, it becomes impossible for an eavesdropper to determine if the photon pair is from the selected subset or the rest, just by looking at the arrival time. So, Alice uses two single mode fibres of different lengths in the two output arms of the BS. Photons that traverse the transmitted arm and become D polarized later have a fixed time delay ($\Delta t$) as compared to photons that traverse the reflected arm and become H polarized. Alice also records the arrival time of each heralding photon detected in her detector using a time tagger (HydraHarp 400) connected to the detector. This is crucial in determining whether the corresponding pair photon is delayed or not, before it is sent to Bob, and hence its polarization state. Thus, by introducing a time delay in the path of one of the photons and knowing the arrival time information of its partner photon (heralding arm), Alice is able to determine the polarization state of the photon that is sent to Bob.\par

In order to remove any distinguishability in the spatial degree of freedom, outputs from both single mode fibres recombine at another 50-50 BS, and only one output arm of the BS is used to send both V and D polarized photons to Bob.\par

\begin{figure}[!ht]
    \centering
    \includegraphics[width=\linewidth]{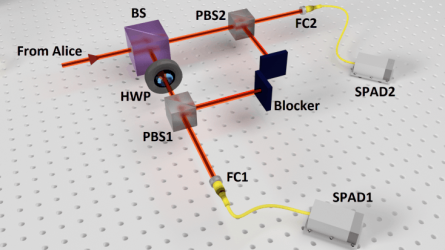}
    \caption{Schematic of Bob's module. \textbf{BS}: 50-50 non polarizing beam splitter; \textbf{HWP}: half-waveplate; \textbf{PBS1, PBS2}: polarizing beam splitters; \textbf{FC1, FC2}: fibre couplers; \textbf{SPAD1, SPAD2}: single photon avalanche detector.}
    \label{fig:B}
\end{figure}

\textit{Bob-substation}: In Bob's part of the experimental architecture as shown in Figure\,\ref{fig:B}, a 50:50 beam splitter (BS) is placed in the path of the incoming photons, where each photon has 50$\%$ probability to go to the transmitted arm and 50$\%$ to go to the reflected arm. In the transmitted arm of the BS, one polarizing beam splitter (PBS2) is placed and a fibre coupler (FC2) collects any photon that transmits through the PBS and sends it to a single photon detector (SPAD2). So, in that arm only D polarized photons have 50$\%$ probability to get detected while V polarized photons have 0$\%$ probability. In the reflected arm of the BS, a similar combination of PBS (PBS1) and fibre coupler (FC1) is placed with an additional half-waveplate (HWP) just behind the PBS. This HWP converts D photons to V photons and vice versa. So, only V photons sent by Alice pass through the PBS and get detected in this arm half the number of times, but no D polarized photon is detected. So, to summarize, any detection in the transmitted arm of the beam splitter definitely means that the photon is D polarized (assigned bit value 0), and any detection in the reflected  arm definitely means the photon is V polarized (assigned bit value 1). Similar to Alice, Bob also records photon time-stamping data from the two detectors at his end.\par


\textit{Post-processing:} In the post processing stage, Bob shares only his time-stamping data publicly, but does not indicate which time-stamping data comes from which detector. Alice compares her time-stamping data with Bob's data and measures time-difference for all detected photon pairs. Let us say the distance between Alice and Bob is $d$, then the time-difference should be ideally $T=d/c$, where $c$ is the speed of light. For the D photons, where Alice applied additional time delay $\Delta t$, time-difference becomes $T+\Delta t$. Alice considers those events where time-difference is closer to $T$ (within a small time window around $T$) as bit value $1$ ( in this case, V polarized photons were sent by Alice to Bob), and considers those events where time-difference is closer to $T+\Delta t$ as bit value $0$ (D polarized photons were sent in this case).\par

Alice then sends back Bob's time stamping data, with the modification that all those events that couldn't make it to Alice's final key are omitted. Based on this, Bob generates his final key. Thus, they both agree on the same generated key.\par

This brings us to an important comment regarding the use of heralded single photons for our implementation and its ramifications. For the purpose of security of the generated key, generation of ideal single photons $\left( \text{ with Fock state }\left|1\right\rangle\right)$ is essential. In case of a multi-photon source, an eavesdropper may apply photon number splitting (PNS) attack. One way to verify the single photon distribution is to look for the anti-bunching property where the probability of generating two consecutive photons within the coherence time is negligible. For this purpose, we can measure the normalized second order coherence or $g^{2}$ by performing a Hanbury, Brown and Twiss (HBT) type experiment. For an ideal single photon source $g^{2}(\tau=0) = 0$, where $\tau$ is the time interval between two consecutive photons.\par

In a real-world experiment, there are stray photons that get detected in Bob’s detection module along with single photons that are sent by Alice. There are also other sources of noise like dark noise of the detector, electrical noise, etc. A single photon detector (SPAD) cannot distinguish noise from the actual signal. These noises increase the quantum bit error rate (QBER) in the generated key. For noise cancellation, heralded single photon source plays an important role. In a heralded photon source, correlated photons are always generated as a pair. Detection of one photon in each pairs ensures the presence of the other photon. Therefore, Alice and Bob post-select only those events where Alice detects one photon and Bob detects the other photon of the same pair (i.e. coincident events) and consider them as part of the signal.\par

\subsection{Data analysis}
\label{subsec:dataanalysis}

In the data processing stage, we measure three important parameters that are the quantifiers of the performance of the QKD experimental setup. These parameters are key rate, quantum-bit-error-rate (QBER) and asymmetry of the key.\par

For key rate, we measure average number of bits in the sifted key (including error bits) generated per second. In B92 protocol, sifting includes post-processing of the data set, where Alice and Bob selects only those events where Bob's detectors show positive outcome and the detection occur within some predefined time window. After the completion of the protocol Alice and Bob both have a key that should be identical in the absence of any channel noise and eavesdropping activity. We measure the number of error bits by comparing each value of the bits for the same bit position in the two keys. The number of error bits divided by the total key length gives the value of the QBER. Asymmetry quantifies the disparity between the number of 0 and 1 bits in the final error-free key shared by Alice and Bob.\par

\begin{figure*}[!ht]
\centering
\includegraphics[scale=0.75]{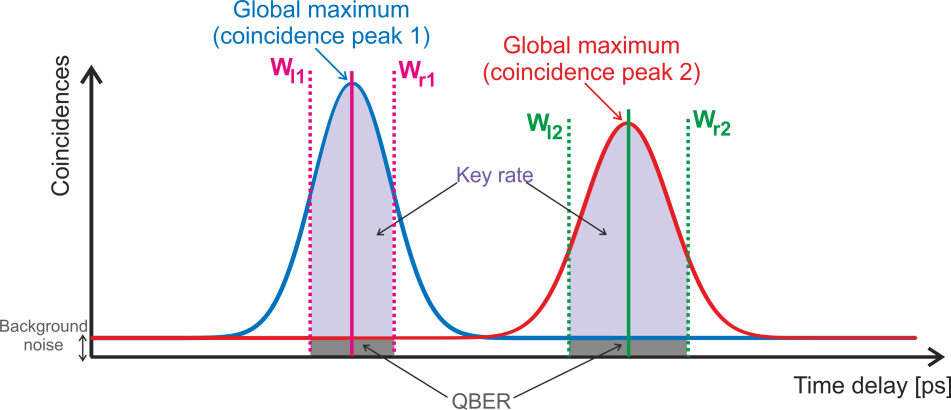}
\caption{A simplified schematic of the output of two independent coincidence detection: Alice and Bob's \textbf{V} basis (coincidence peak 1 in blue) versus Alice and Bob's \textbf{+} basis (coincidence peak 2 in red). The background noise zone indicated below the flat portions of the blue and red curve represents the unwanted coincidence detections from stray light sources and dark noise of the photodetector. The symbols $W_{\text{l1(2)}}$ and $W_{\text{r1(2)}}$ represents the left and right markers of the time window around the maximal coincidence point for coincidence peak-1 (2). The total coincidences within the chosen window around the central maximum from both curves contribute to the \enquote{key rate} (or signal - marked in purple); while those within the background noise zone contribute to the \enquote{QBER} (or noise - marked in grey). Note that in reality the coincidence curves are not typically smooth functions and contain a lot of kinks (local optimal points) around a central global maximum.}
\label{fig:schematic}
\end{figure*}

In order to measure key length, QBER, and asymmetry from each data set, we have applied two types of optimization methods, namely A \& B, on every data set. In both the methodologies, at the beginning Alice and Bob's recorded time stamping data are compared and plotted as a function of time difference between the two. The schematic in Figure\,\ref{fig:schematic} shows two distinct coincidence peaks due to the time delay of around 10 ns introduced in Alice's setup. First peak (blue) represents those coincidence events where Alice sent V photon and Bob measured correctly. Second peak (red) represents those coincidence events where Alice sent D (delayed) photon and Bob measured correctly. The extension of the red curve under the blue curve represents those events when Alice sent V, but Bob measured it wrongly as D; due to the noise introduced by optical components as well as the transmission channel. Similarly, the extension of the blue curve under the red peak represents those events when Alice sent D, but Bob measured it wrongly as V. We fix some time window around both peaks ($\text{W}_{\text{l1}}$ to $\text{W}_{\text{r1}}$, and $\text{W}_{\text{l2}}$ to $\text{W}_{\text{r2}}$) and measure the area under the curves. The sum of the total area under the blue and the red curve for their consecutive time windows gives the value of the key length. The sum of the total area under the red curve from $\text{W}_{\text{l1}}$ to $\text{W}_{\text{r1}}$, and the blue curve from $\text{W}_{\text{l2}}$ to $W_{\text{r2}}$ gives the number of total error bits (that contribute to the QBER).\par


Nevertheless, ideally for secure key generation the probability for obtaining  any key string of $N$ key bits among the $2^N$ set of possible key strings should be equal; i.e. any key can be generated with the probability of $\frac{1}{2^N}$. If this ideal case is to realized then all the optical components should behave perfectly, i.e. symmetric beam splitter has exactly 50$\%$ probability of both transmission and reflection, both coincidence peaks should be identical. However, in real experimental scenario the two peaks may be different. In other words, we generated a large number of key strings and found that all of them have equal asymmetry of around 60:40 due to the asymmetric beam-splitting,  i.e. our device itself introduces the bias. If the device would have been generating perfectly random outputs, we would have obtained an 
distribution peaked at 50:50 rather than a fixed asymmetry value.\par

In order to counter this issue, we have implemented two different optimization strategies: \enquote{A} and \enquote{B} (refer to Appendix\,B for the detailed procedure of the two strategies). In strategy A, we optimize each key string individually and obtain a symmetry of nearly 50:50 among them. More specifically, the whole purpose of this optimization method is to find the value of the two coincidence time windows or the position of $\text{W}_{\text{l1}}, \text{W}_{\text{r1}}, \text{W}_{\text{l2}}$, and $\text{W}_{\text{r2}}$, such that the QBER remains below the threshold value (4.8$\%$) \cite{PhysRevA.72.012332} and the key is symmetric (the number of 0's and 1's in the key string is equal). However, this approach is still insecure as the asymmetry distribution now becomes fixed at 50:50 instead of 60:40. Moreover, this step requires deletion of some bit values, that lowers the final key length. Nevertheless, this strategy drastically reduces eavesdropper’s ability to extract additional information about the key.\par 

In our second strategy, B, while maintaining the same length of the two time-windows, we changed their positions such that we obtain the maximum key length within the QBER bound, for each key string separately. The key length value was found to increase in this process, but the asymmetry value remained fixed at 60:40. However, it is possible in principle, to adapt a modified QBER analysis method such that eavesdropper’s additional information gain due to any biases can be monitored.

In order to report the average key rate, we run the protocol for 10 seconds and repeat the same for 20 iterations. The final key rate is then averaged over the 20 key length values. In order to ensure that we choose a run-time of the protocol such that the standard deviation (SD) by mean (M) i.e. $SD/M$ of the reported average key rate is very small, we collected data continuously for a longer time duration (say, 100 seconds), applied the bootstrapping technique as discussed in Appendix A, 
and obtained a $SD/M$ plot (refer Figure\,\ref{fig:bootstarp_result-3}) as a function of the runtime for fixed number of iterations. We found $SD/M$ for 10 seconds run-time and 20 iterations to be 0.016$\%$.

\section{Simulation toolkit}
In this section, we will discuss the principle aspects and the current stage of the implementation of the toolkit. Following an overview, we go on to discuss the various assumptions that have been considered while simulating the experimental demonstration. In the later sub-sections, we provide a detailed discussion on the various modules that comprises the toolkit.
\subsection{Overview}
\begin{figure}[!ht]
    \centering
    \includegraphics[width=\linewidth]{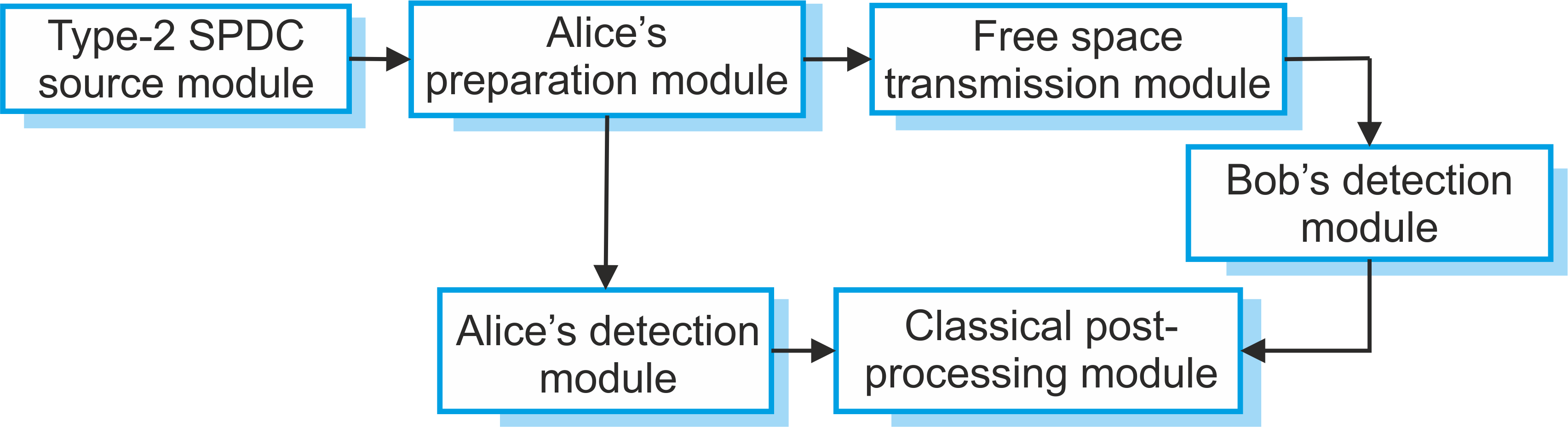}
    \caption{B92 simulation structure}
    \label{fig:b92_sim}
\end{figure}

 We have simulated the experimental demonstration of the B92 protocol discussed in Section\,\ref{sec:experiment} using the simulation toolkit 'qkdSim'  discussed in Section\,\ref{sec:framework}. Different modules have been developed for the respective choices corresponding to the source, detection and transmission components used in the actual experiment. The Table\,\ref{tab:inputs} lists down the various choices for the general inputs to the toolkit and the Figure\,\ref{fig:b92_sim} shows the interconnection of the various modules developed for the toolkit.\par
\begin{table}[h!]
  \centering
  \begin{tabularx}{8cm}{Sl|X}
    \textbf{Choices} & \textbf{Inputs} \\
    \hlineB{1.2}
    Type of protocol & B92 QKD protocol \\
    Type of source & {Type II collinear degenrate SPDC source} \\
    {Type of transmission channel} & Free-space \\
    Distance of transmission &  2 metres \\
    Type of detection & {fibre-based detectors and TCSPCM} \\
    Protocol run-time & {20 runs for 1 second each} \\
    Security parameter & QBER threshold \\
    Environmental conditions & {In-lab (daytime and nighttime)}
  \end{tabularx}
  \caption{Choice of inputs}
  \label{tab:inputs}
\end{table}

The modules are interconnected for the flow of logic and mimics the path of the photons in the actual experimental setup. The output from the source module, based on type -II collinear degenerate SPDC process, is passed to Alice's preparation module. The signal states are encoded for transmission and relayed to the transmission module while the heralding states are passed to Alice's detection module. The transmission module simulates the transmission of single photons over the quantum channel between Alice and Bob and the detection of the received signal states is simulated in Bob's detection module. The outputs from Alice's and Bob's detection modules are fed into the classical post processing module, which then gives as output the sifted key rate, QBER and the key symmetry.\par

Each of the modules is constructed with the help of various sub-modules corresponding to the various physical components and processes that play important roles in the experimental implementation of the protocol. The inputs to the modules can be categorized as user inputs, set parameters and outputs obtained from preceding modules. The user inputs refers to certain choices made by the user whereas the set parameters refers to the specification of the various components that the setup consists of. Before we discuss in detail the structure and working of each of the modules, we discuss the various assumptions that have been considered for the simulation, in the following sub-section. \par
\subsection{Assumptions}
\label{subsec:assumptions}
Though the simulation toolkit takes into account various non-ideal aspects of experimental implementation of a QKD protocol, it is not exhaustive. Thus it is essential to provide a detailed list of assumptions that have been considered while implementing the architecture at each stage of the iteration. The assumptions considered for the simulation of the B92 protocol are listed as follows.
\begin{enumerate}
\item The time stamps associated with each of the photon pairs are evolved through the experimental setup comprising different optical components. However, in the sections of the experimental setup where the photon pairs travel identical path lengths or the possible paths that can be traversed by a single photon have identical length, the time taken to travel is not accounted. For example, the detectors in Bob's detection module are assumed to have been positioned equidistant from the BS (Figure\ref{fig:B}). 
\item The pump laser output is assumed to be a symmetric Gaussian beam. The generated photons are also considered to have similar beam properties properties as that of the pump laser. Thus each photon is associated with a Gaussian distribution for the intensity. 
\item It is assumed that the alignment of the optical and mechanical components in the experimental set up is achieved up to maximum precision, limited only to the error introduced by the least count of the screws of the mounting components and stages. 
\item In the simulation of the type-II collinear degenerate SPDC process, the frequency linewidth of the pump laser is assumed to be extremely narrow and hence, the frequency distribution of the signal and idler photons is not taken into account. Additionally it has also been assumed that the pair generation takes place only at the centre of the crystal and the crystal medium is lossless.  
\item In the experiment, it has been observed that the important parameters of the system such as the laser power, temperature of the crystal etc. does not fluctuates significantly over the period of time for the data acquisition. Thus, in simulation, we have the assumed the system to be time-invariant and all the parameters have been assumed to be constant over the run-time of the simulation.
\item Any effect in the phase or polarization of the photons due to the transmission over free-space and optical fibre channel is neglected. 
\item No eavesdropping strategy or attack has been considered for the simulation. A security parameter that corresponds to a threshold QBER derived based on the protocol and independent to the simulation is taken as input to the system.
\end{enumerate}

We have discussed the various assumptions that have been considered in the implementation of the simulation toolkit and now we go on to discuss each of the modules shown in Figure\,\ref{fig:b92_sim} in order of the path followed by the photons in the actual experimental setup. The source module is discussed first followed by the modules corresponding to the preparation, transmission and detection of the signal photons and concluding with the post-processing module. For each of the modules, a brief introduction is followed by a brief discussion on the inputs and outputs to the modules, the module structure and the algorithmic overview. Each modules are constructed using relevant sub-modules and to avoid redundancy, the detailed discussions on the various sub-modules are given in Section\,\ref{sec:process} and \ref{sec:components}.

\subsection{Type-II SPDC source module}
\label{sec: source}
\begin{figure*}[!ht]
\centering
\includegraphics{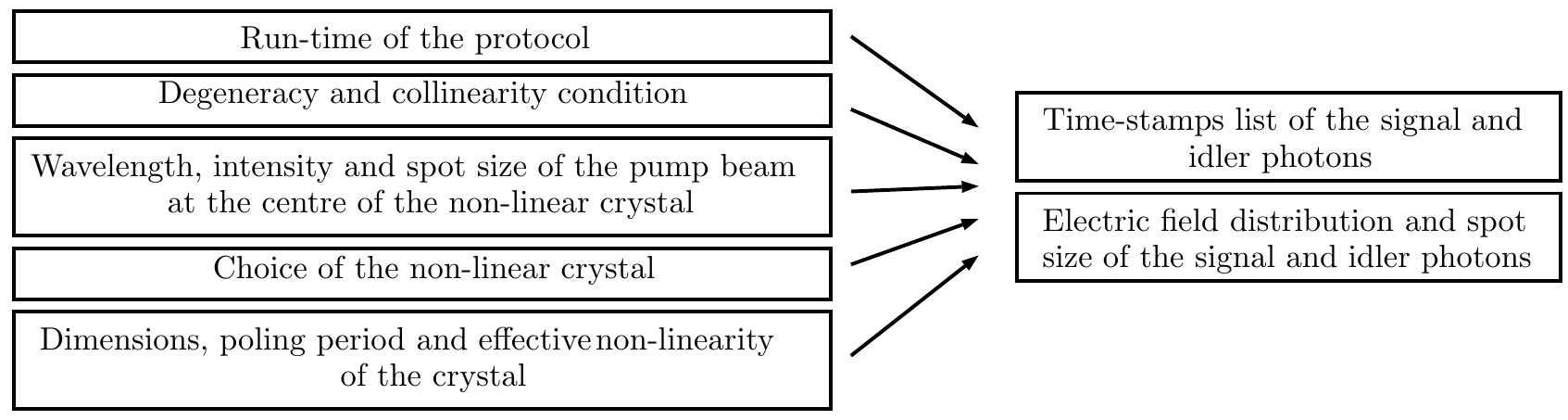}
\caption{Source module overview}
\label{fig:source_tab}
\end{figure*}

The type -II SPDC source module simulates the generation of photon pairs in a type-II SPDC process and the temporal distribution of the photon pair generation events at the crystal. At present, the module simulates the specific case of type-II collinear degenerate SPDC process by quasi phase matching (QPM) using a ppKTP crystal. The inputs and the structure of the module are in accordance with this specific case and the sub-modules are also developed accordingly.
Figure\,\ref{fig:source_tab} lists down the user inputs to the source module, the set parameters and the outputs of the module. [S] denotes set parameter.

\begin{figure}[!ht]
    \centering
    \includegraphics[width=\linewidth]{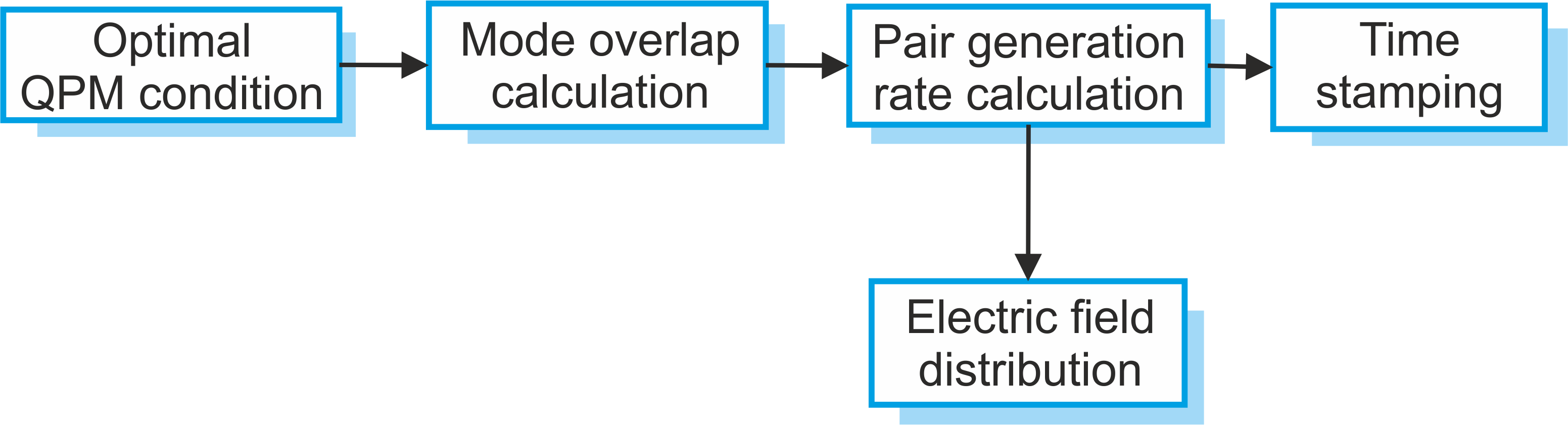}
    \caption{Type-II SPDC source module layout}
    \label{fig:source}
\end{figure}

With respect to the structure, the source module is constructed of sub-modules for simulating the various aspects of the SPDC process and the layout is shown in Figure\,\ref{fig:source}. 

The optimal QPM condition sub-module takes as input the pump wavelength and the poling period of the crystal and calculates the phase matching temperature for the degenerate condition. The mode overlap function sub-module then calculates the pair generation probability per pump photon for the given conditions of pump beam characteristics, crystal dimensions and temperature for the phase matching condition. The total pair generation rate is calculated by taking into account the pump beam intensity at the crystal and the simulated pair generation rate obtained per pump photon from the preceding sub-modules. \par

The time stamp generation sub-module takes as input the pair generation rate and the total run-time of the protocol and creates a list of time stamps of the event of generation of photon pairs at the crystal in respect to a global clock. The spot size of the signal and idler photons is calculated from the pump beam spot size and the electric field distribution sub-module generates a list of electric field amplitudes, at different points along an axis perpendicular to the direction of propagation of photons, that follows a Gaussian distribution. Separate lists of time stamps for the signal and idler photons along with the list Gaussian distribution of their electric field are returned to the main module of the toolkit.\par

\subsection{Alice's preparation module}

\begin{figure*}[!ht]
\centering
\includegraphics{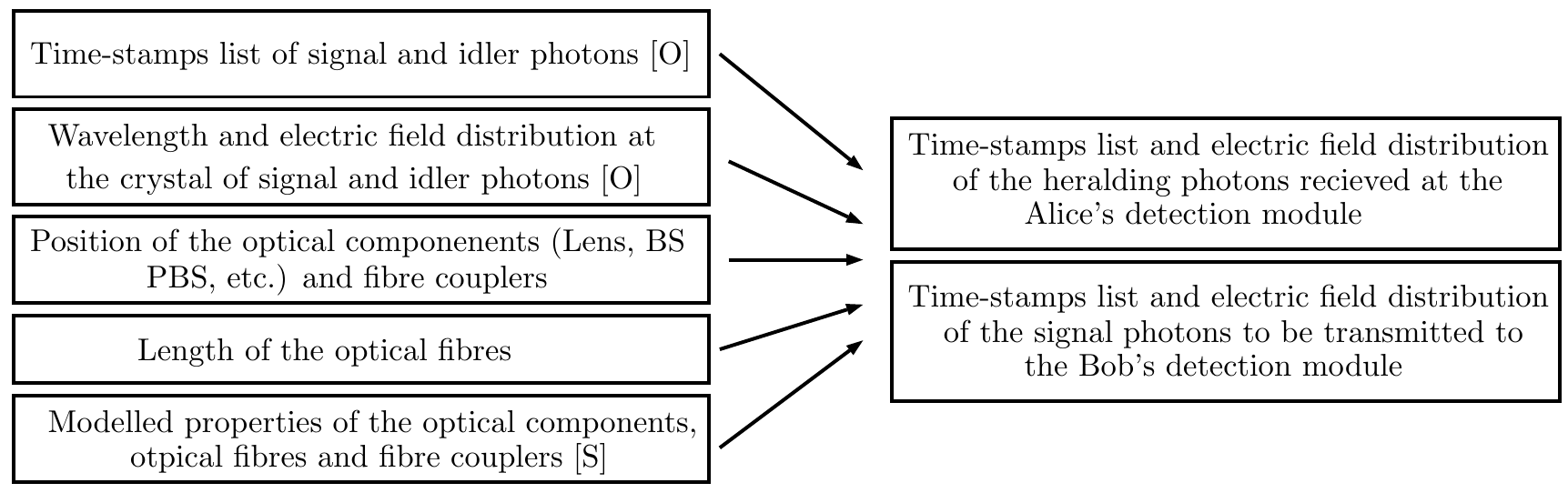}
\caption{Alice's preparation module overview}
\label{fig:prep_tab}
\end{figure*}

Alice's preparation module simulates the stage where Alice prepares the states that are to be sent to Bob over the quantum communication channel. The signal photons are encoded for transmission to Bob whereas the idler photons are transmitted to the detection component of Alice.
Figure\,\ref{fig:prep_tab} lists down the user inputs to the source module, the set parameters and the outputs of module. [S] denotes set parameter and [O] denotes inputs that are given as output by the previous module(s).\par
\begin{figure*}[!ht]
\centering
\includegraphics[scale=0.45]{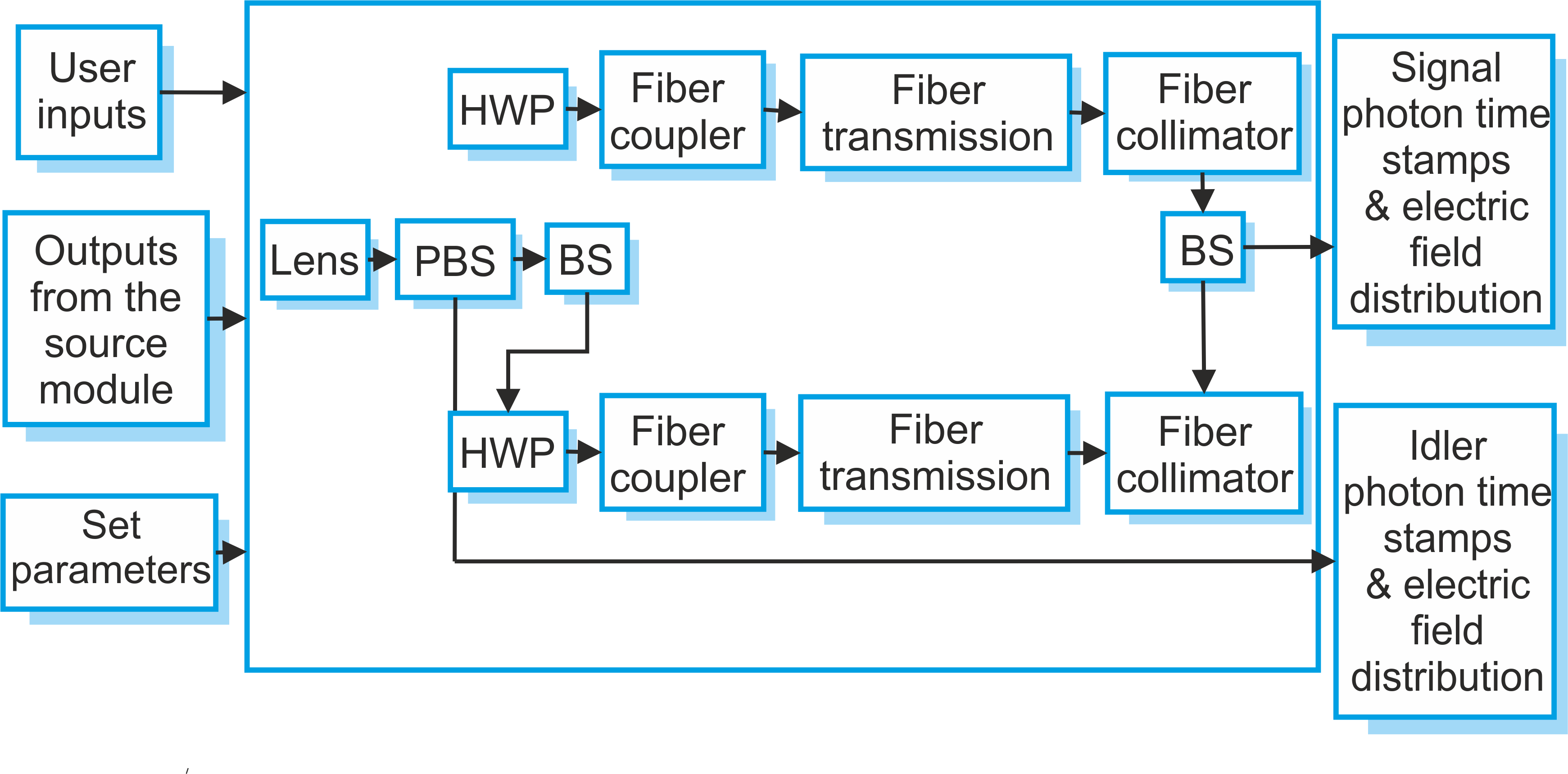}
\caption{Alice's preparation module layout}
\label{fig:prep}
\end{figure*}

Alice's preparation module is constructed of various sub-modules of different physical components and processes to process the time stamps and the electric field distribution of the generated photon pairs from the type-II SPDC source module. The structure of the module is shown in Figure\,\ref{fig:prep}. The arrows refers to the flow of logic within the module and the two separate outputs are generated corresponding to the signal and the heralding arm of Alice.

The initial section of the module is common to both the stream of photon pairs generated from the crystal (signal and idler). The time stamps and electric field distribution arrays are first evolved through the lens sub-module that provides loss and change of phase to the incident photons. The resultant electric field distribution of the photons at the signal and the heralding end of the module are calculated using the input electric field distribution array. The time stamps arrays are further evolved with the other sub-modules. The PBS sub-module generates the signal and heralding photons time stamps arrays and while the former is passed to the following sub-modules, the later is not.\par
The signal time stamps array is separated into two with each of the arrays being further evolved through the HWP, fibre coupler, fibre transmission and the fibre collimator sub-modules in order. At the HWP module, the photons are projected to specific polarization desired for transmission and encoded with a bit-value ('0' or '1') corresponding to the polarization. Each element of the time stamps lists of the photons is appended with the corresponding bit-value. The resultant arrays from the two different streams are further merged at the BS sub-module and the final array is generated and stored as the signal time stamps array. The resultant arrays are returned to the main module of the toolkit.

\subsection{Transmission module}
\begin{figure*}[!ht]
\centering
\includegraphics[width=\linewidth]{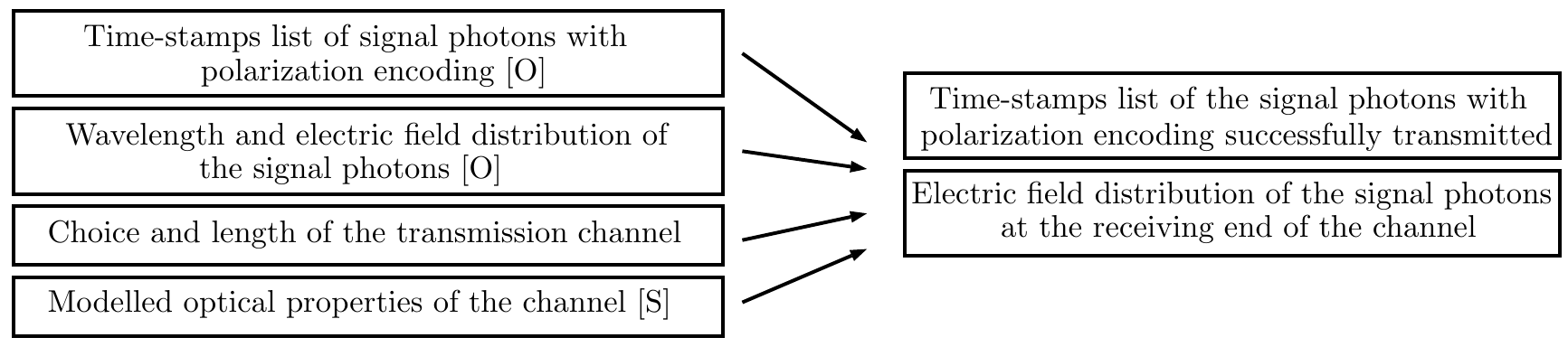}
\caption{Transmission module overview}
\label{fig:trans_tab}
\end{figure*}

The transmission module simulates the transmission of the photons sent by Alice to Bob that can be over a free-space or fibre-based channel. Figure\,\ref{fig:trans_tab} lists down the user inputs to the source module, the set parameters and the outputs of module. [S] denotes set parameter and [O] denotes inputs that are given as output by the previous module(s). It is important to note that the choice of the channel that forms a general input to the system is specifically used in the transmission module and in accordance with the choice, the respective sub-module corresponding to the free-space or optical fibre-based transmission is used. The transmission module incorporates the in-lab free-space transmission sub-module to simulate the transmission of the single photons from Alice to Bob that is in accordance with the experimental setup been simulated. \par
\begin{figure}[!ht]
    \centering
    \includegraphics[width=\linewidth]{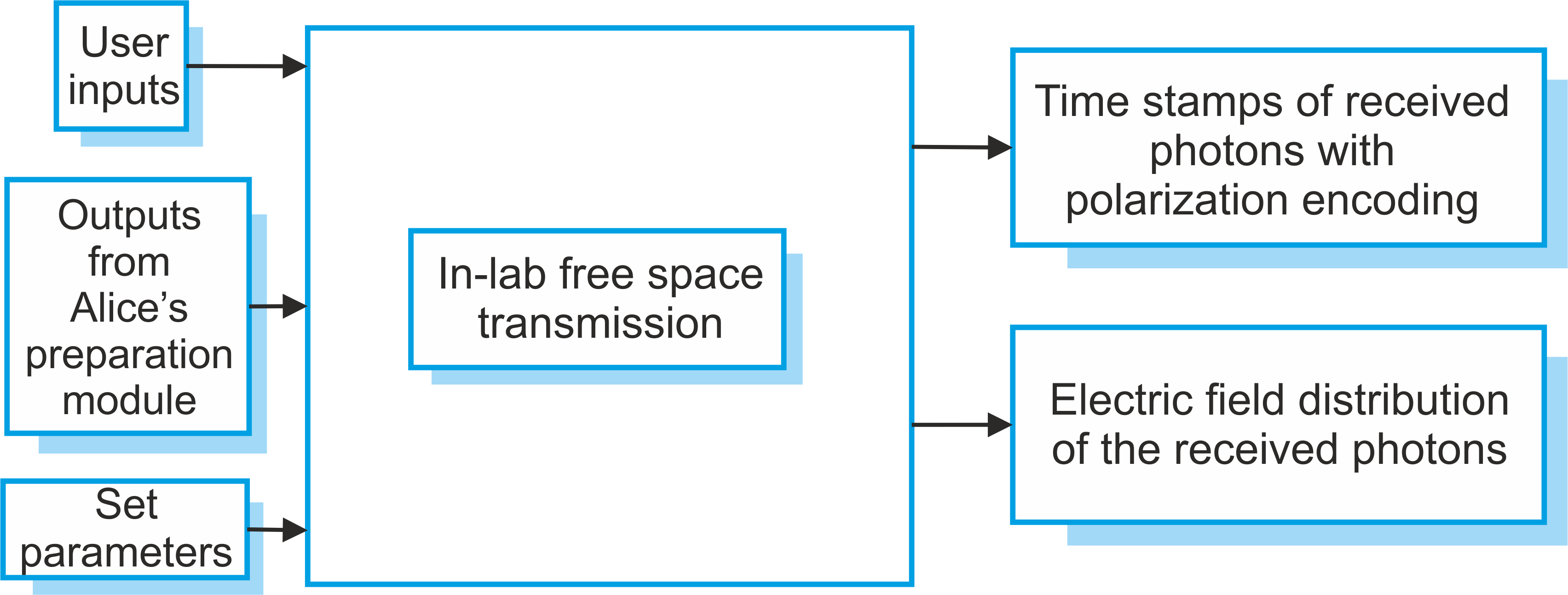}
    \caption{Transmission module layout}
    \label{fig:trans}
\end{figure}
The time stamps of the transmitted signal photons and the electric field distribution are evolved with the in-lab free-space transmission module as shown in Figure\,\ref{fig:trans}. The losses in the channel incurred by the photons are accounted and the time stamps of the photons received at Bob's detection module are returned to the main module along with polarization encoding and electric field distribution of the received photons.

\subsection{Bob's Detection module}
\begin{figure*}[!ht]
\centering
\includegraphics[width=\linewidth]{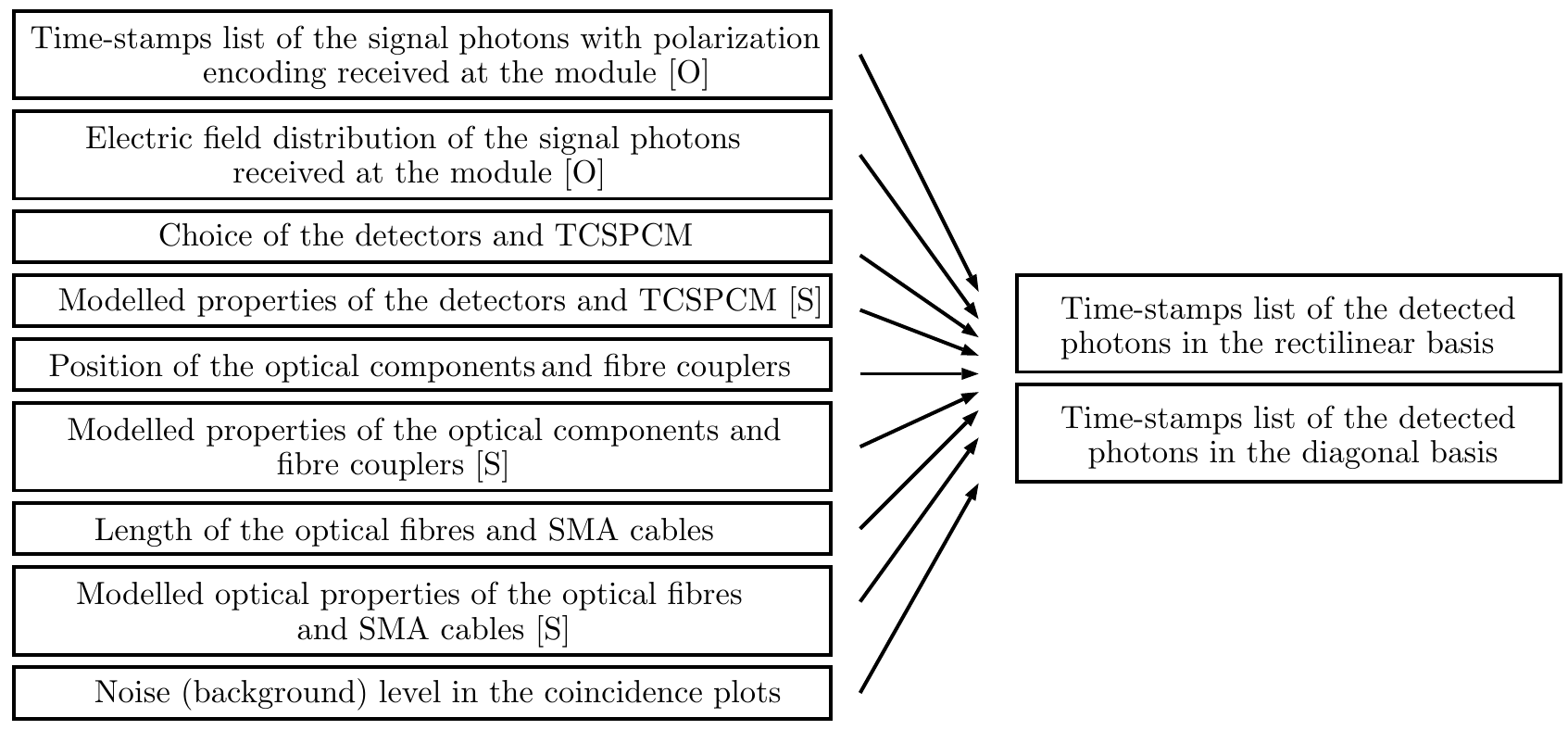}
\caption{Bob's detection module overview}
\label{fig:bobd_tab}
\end{figure*}

\begin{figure*}[!ht]
\centering
\includegraphics[scale=0.42]{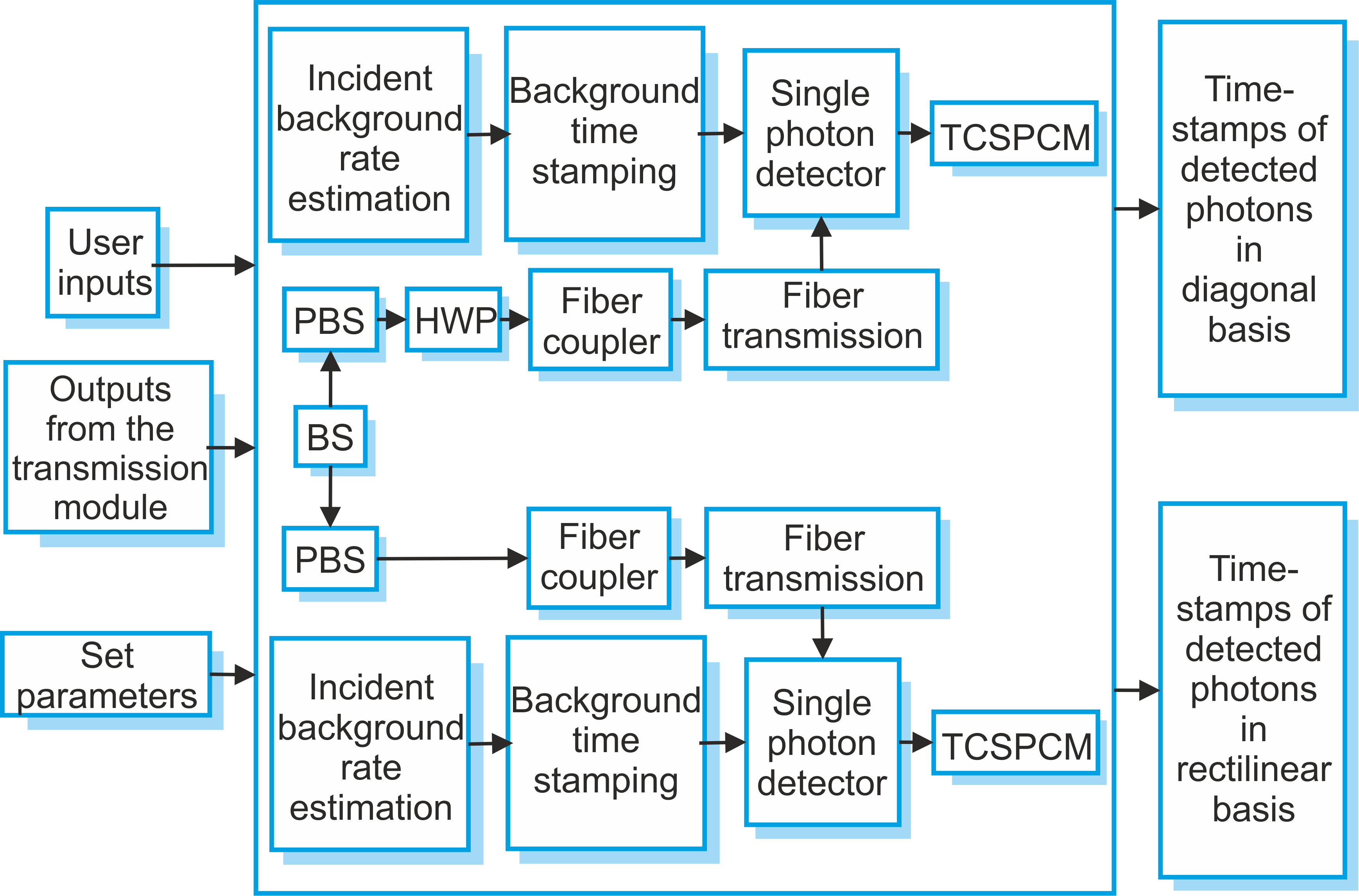}
\caption{Bob's detection module layout}
\label{fig:bobd}
\end{figure*}

This module simulates the detection of the signal photons transmitted by Alice over the quantum channel.
Figure\,\ref{fig:bobd_tab} lists down the user inputs to the source module, the set parameters and the outputs of module. [S] denotes set parameter and [O] denotes inputs that are given as output by the previous module(s). The choice of the detection components i.e. the type of the single photon detector and the time correlated single photon counting module (TCSPCM) dictates the use of respective sub-modules within this module.
Bob's detection module is constructed of the detection components chosen by the user as well as the requirements based on the protocol. As per the experimental set up, the sub-modules for fibre-based single photon detectors and TCSPCM have been used. The structure of the module is depicted in Figure\,\ref{fig:bobd}. The arrow denotes the flow of logic among the different sub-modules. The detection of photon in the rectilinear and diagonal basis are simulated separately with the respective inputs corresponding to the noise level.
The time stamps of the received photons along with the polarization encoding are evolved through the BS sub-module which splits the received time stamps into two separate arrays corresponding to the random basis choice in that the photons will be measured. For the rectilinear basis, the photons are evolved with the fibre coupler and fibre transmission sub-modules to obtain the photons that are received at the detector. In parallel, the electric field distribution of the received photons are evolved according to the distance of the fibre-couplers and the coupling efficiency is obtained using the fibre coupler sub-module. Similar logic is followed for the diagonal basis with an addition of the HWP sub-module that enables the simulation of the basis projection.
From the sub-modules related to the background detection, the time stamps list of the background photons are obtained and merged with the time stamps of the signal photons received at the detector. At this point, the polarization information of the received photons is deleted as the information is redundant once the photons have been received at the detectors. The time stamps of the photons are then evolved with the single photon detector and TCSPCM sub-module to finally generate the time stamps list of the detected photons. A bit value corresponding to the basis at which the photons are detected are appended to the list of the detected time stamps for each of the photons. This list is then returned to the main module of the simulation toolkit.

\subsection{Alice's Detection Module}
\begin{figure*}[!ht]
\centering
\includegraphics{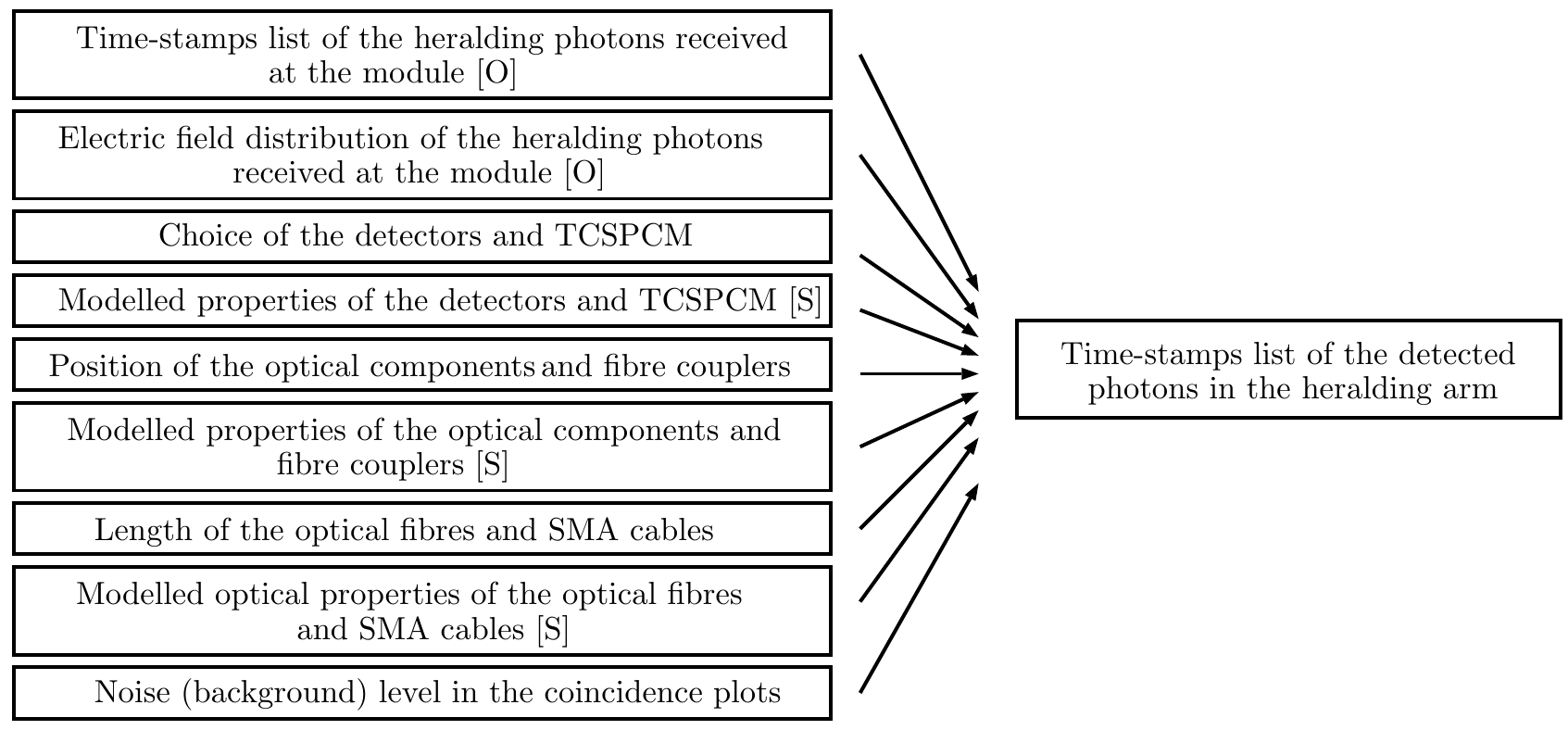}
\caption{Alice's detection module overview}
\label{fig:aliced_tab}
\end{figure*}

Alice's detection module simulates the detection of the heralding photons that are separated from the photon pairs at Alice's preparation module. Figure\,\ref{fig:aliced_tab} lists down the user inputs to the source module, the set parameters and the outputs of module. [S] denotes set parameter and [O] denotes inputs that are given as output by the previous module(s). Similar to Bob's detection module, respective sub-modules corresponding to the choice of the detection components provided by the user are used in this module.

\begin{figure*}[!ht]
\centering
\includegraphics[scale=0.45]{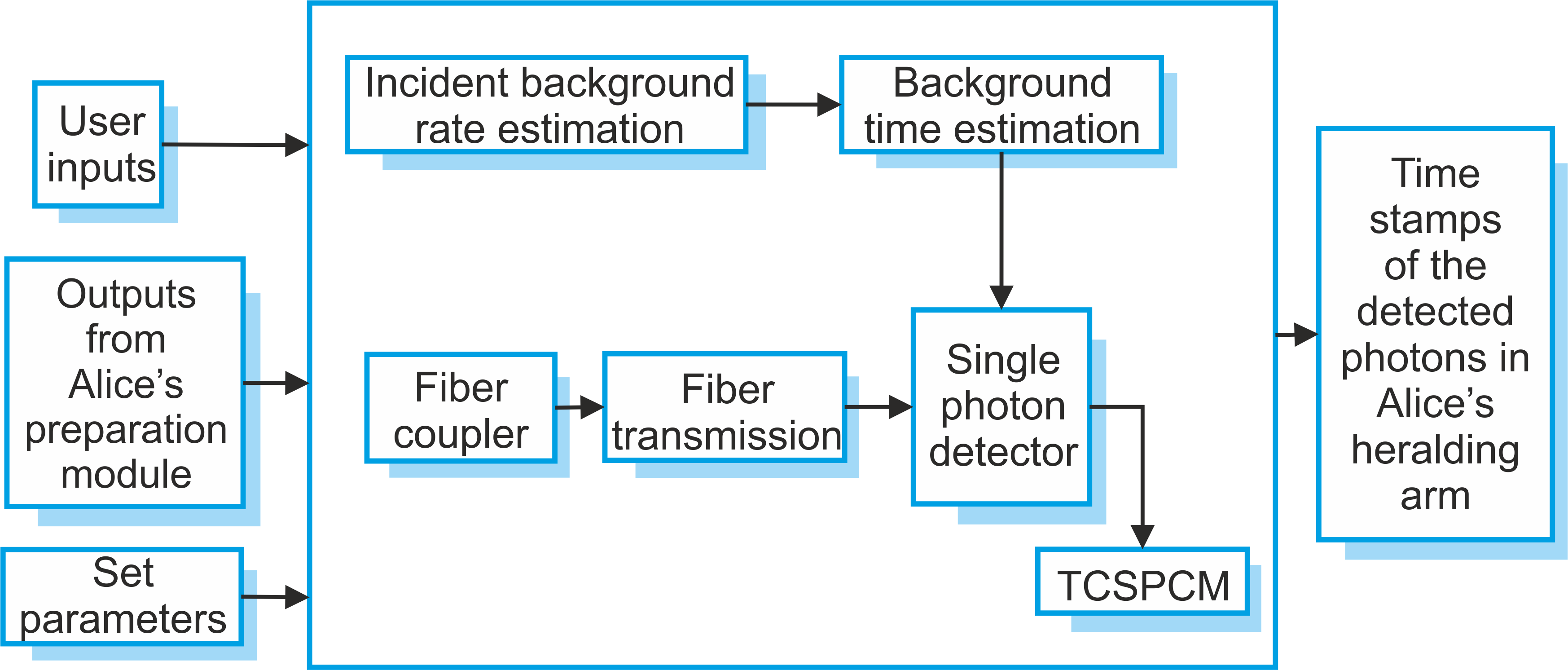}
\caption{Alice's detection module layout}
\label{fig:aliced}
\end{figure*}

Alice's detection module is structured based on the protocol and the implementation that is being simulated. Fibre-based single photon detectors and TCSPCM sub-modules have been used to construct the detection module. The structure of the module is depicted in the Figure
~\ref{fig:aliced}. The time stamps and the electric field distribution of the heralding photons at the fibre coupler position at the heralding arm of Alice are evolved through the sub-modules in series. The fibre coupler sub-module uses the field distribution to simulate the coupling efficiency while the fibre transmission sub-module generates the time stamps of the photons received at the detector. Similar to Bob's detection module, the incident background rate is estimated and time stamps are generated with the help of the respective sub-modules and are merged with the time stamps of the heralded photons received at the detector. The detection is then simulated with the single photon detector and the TCSPCM sub-module and the list of time stamps of the detected heralding photons is generated and returned to the main program.
\subsection{Classical post-processing module}
\begin{figure*}[!ht]
\centering
\includegraphics{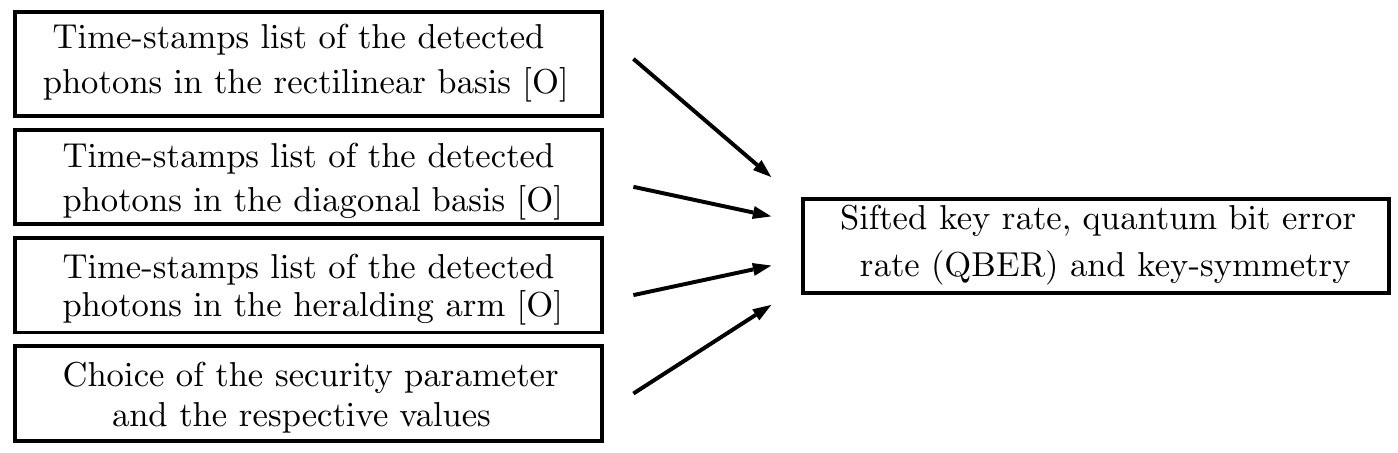}
\caption{Classical post processing module overview}
\label{fig:cpp_tab}
\end{figure*}

\begin{figure}[!ht]
    \centering
    \includegraphics[width=\linewidth]{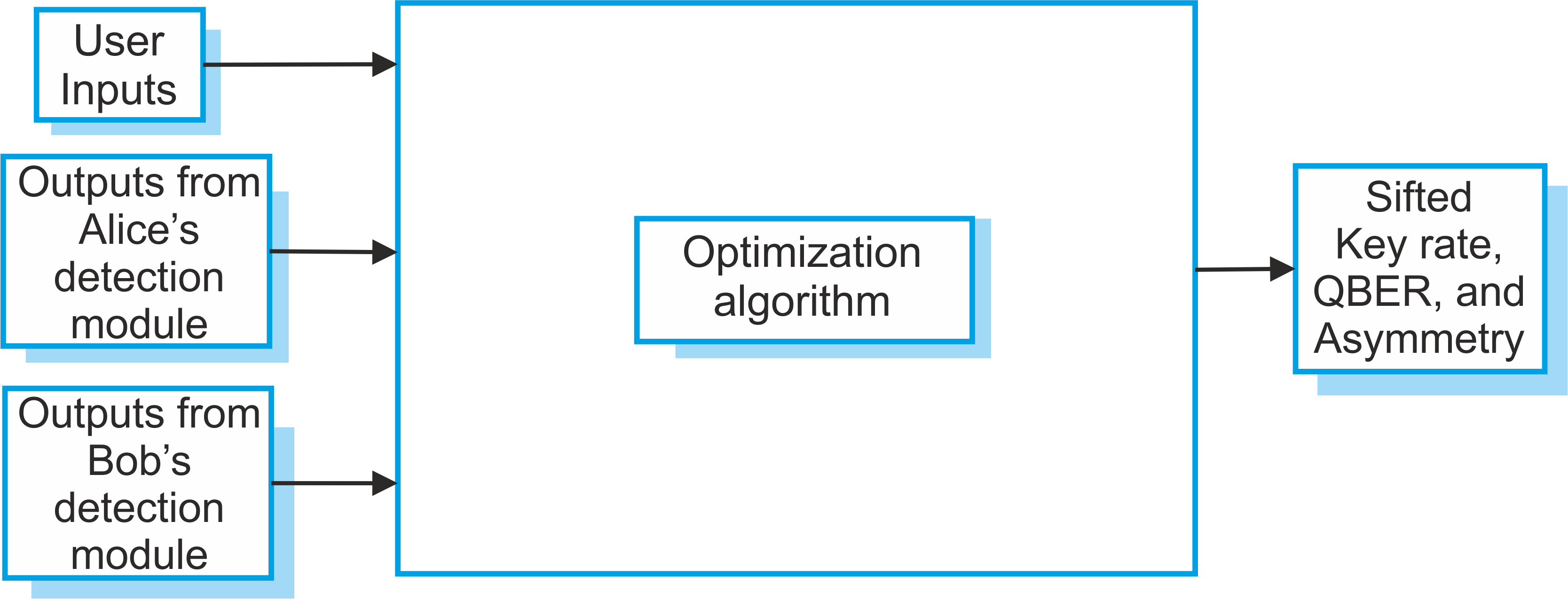}
    \caption{Classical post-processing module layout}
    \label{fig:cpp}
\end{figure}

The classical post-processing module simulates the post-processing of the data after the execution of the protocol. The Figure\,\ref{fig:cpp_tab} lists down the inputs to the classical post processing module and the outputs of the module. [O] denotes inputs that are given as output by the previous module(s).
The module incorporates the optimization strategies that have been developed for implementation of the B92 QKD protocol based on single photon sources.
Depending on the user choice for the security parameter, the corresponding optimization algorithm sub-module is used. Figure\,\ref{fig:cpp} depicts the structure of the module.

\section{Modelling physical processes}
\label{sec:process}
In the current implementation of the QKD simulation architecture, various physical processes have been simulated as listed in Table\,\ref{tab:process}. The physical processes include generation of photon pairs in SPDC process, time stamping of generated photon pairs as well as background thermal photons, propagation of photons and fibre coupling and collimation of the same. In this section, the simulation techniques used for these processes will be discussed in detail.

\subsection{Spontaneous parametric down conversion (SPDC) based source}

\subsubsection*{Background}

Spontaneous parametric down conversion (SPDC) refers to a process of amplification of vacuum uncertainties (or fluctuations) of the optical field in the low gain regime\,\cite{boyd}. In a SPDC process, a photon from the pump (p) laser beam incident on a nonlinear (type-II) crystal such as BBO, PPKTP etc. can originate two other photons: signal (s) and idler (i)\,\cite{mandel-wolf}; as shown in schematic Figure\,\ref{fig:schematic-spdc}. \par
\begin{center}
\begin{figure}[H]
\centering
\includegraphics[scale=0.75]{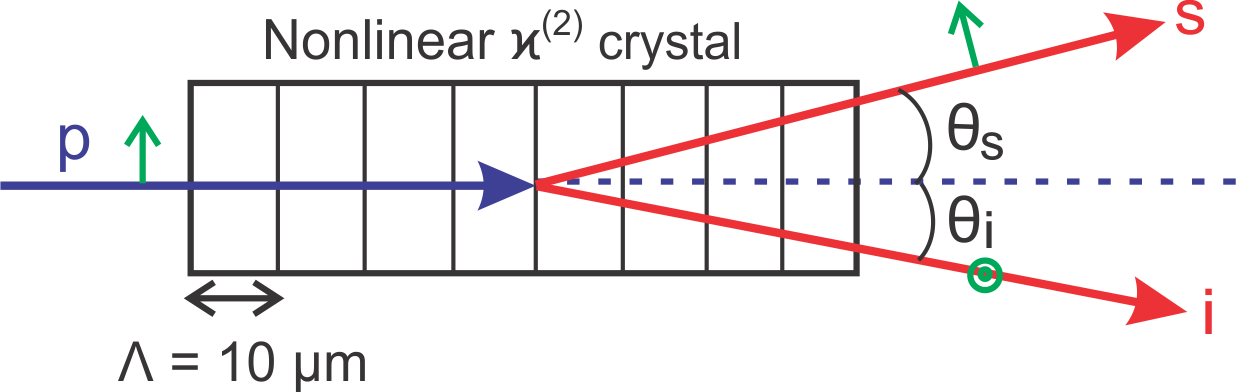}
\caption{A schematic of the SPDC process through a PPKTP crystal of poling period $\left(\Lambda=10\,\mu m\right)$. In SPDC, the input pump (\textbf{p}) photon at 405 nm (in blue) undergoes frequency down conversion and outputs two near-infrared photons (signal (\textbf{s}) \& idler (\textbf{i}), highlighted in red) at double its wavelength (i.e. 810 nm). The green colored dot/arrows represent orthogonal polarization directions.}
\label{fig:schematic-spdc}
\end{figure}
\par\end{center}

\begin{figure}[H]
\centering
\includegraphics[scale=0.75]{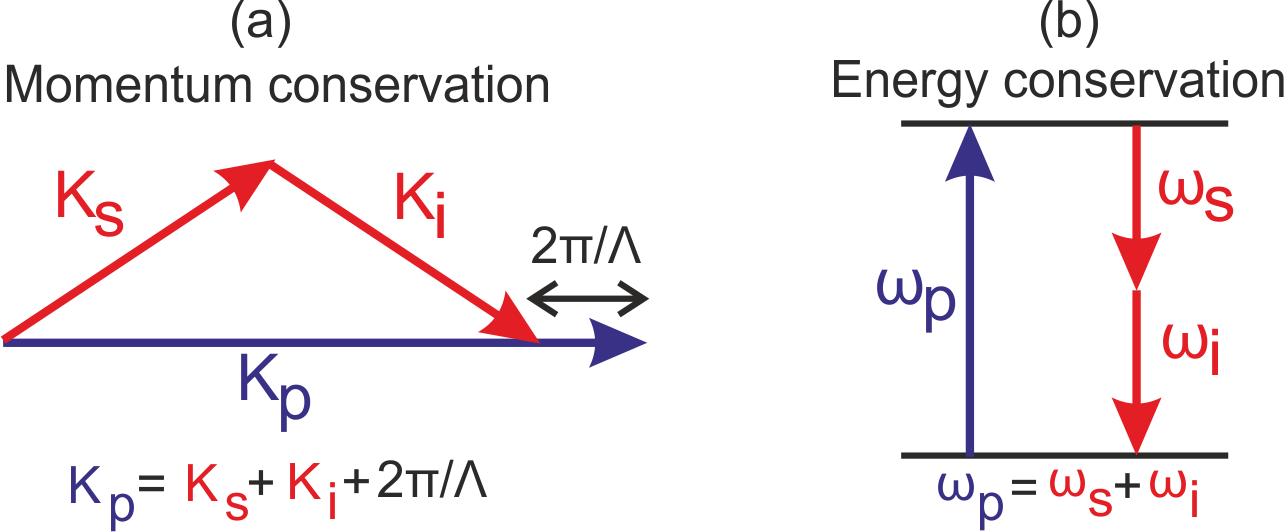}
\caption{Schematic describing the relations for the law of conservation of momentum (left) and energy (right); where $\Lambda=10\,\mu m$.}
\label{fig:spdc-conservation}
\end{figure}

Given that the index of refraction changes with frequency, only certain triplets of frequencies will be phase-matched such that law of conservation of momentum (refer Figure\,\ref{fig:spdc-conservation}(a)) and energy (refer Figure\,\ref{fig:spdc-conservation}(b)) are satisfied. In order to achieve phase matching through the use of a birefringent crystals,
the highest frequency wave $\omega_{p}=\omega_{s}+\omega_{i}$ is polarized in the direction that gives it a lower of the two possible refractive indices\,\cite{boyd}. For the type-II crystal, this choice corresponds to the extraordinary polarization\,\cite{boyd}. Also, the polarization of the pump photon should be the same (extraordinary:
e) as the signal, while the idler should have orthogonal (ordinary: o) polarization (refer the green encircles/arrows in Figure\,\ref{fig:schematic-spdc}).
Thus, for type II crystals, $e_{p}=e_{s}+o_{i}$. \par

From the above considerations for the conservation of energy and momentum in a SPDC process, the phase matching condition for a periodically poled KTP crystal can be obtained by solving the Sellmeier equations using the values of the constants given in \cite{kato2002,bierlein1989,Fradkin1999}. Numerically, the phase matching temperature was calculated to be $44.4^{\circ}C$ considering a pump wavelength of 405 nm along with signal and idler wavelengths of 810 nm (please refer Appendix D for detailed expressions). \par

\subsubsection*{Pair generation probability \& pair generation rate}

\begin{figure}[!ht]
\centering
\includegraphics[scale=0.45]{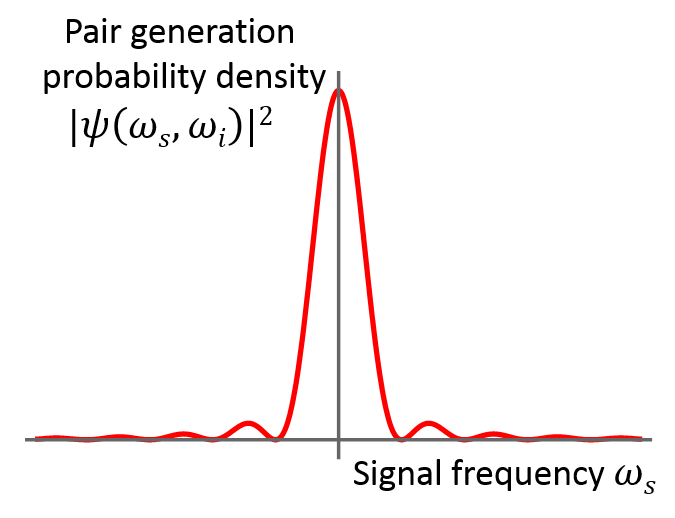}
\caption{Schematic describing the $\textrm{sin\ensuremath{c^{2}}}$
nature of pair generation probability density corresponding to a spectrum
of mode frequencies.} 
\label{fig:sinc2schem}
\end{figure}

The pair generation probability density is the square modulus of the probability amplitude of the SPDC process i.e. $\left|\psi\left(\omega_{s},\,\omega_{i}\right)\right|^{2}$. Now since $\omega_{p}=$$\omega_{s}+\omega_{i}$ and $\omega_{p}$ is the coherent state of the laser source, if we numerically integrate $\psi\left(\omega_{s},\,\omega_{i}\right)$ over a possible range of signal frequencies $\left(\omega_{s}\right)$ then we would obtain a $\textrm{sin\ensuremath{c^{2}}}$ nature plot for this pair generation probability density plotted over a spectrum of signal mode frequencies as illustrated through a schematic in Figure\,\ref{fig:sinc2schem}.

The maximum of this probability distribution provides the corresponding wavelength information at which SPDC pair generation rate would be maximal. The maximum value of pair generation probability obtained numerically can be directly verified with the experimental data. Consequently, the pair generation rate is given by,

\begin{equation}
R_{T}=\left(\int \left|\psi\left(\omega_{s},\,\omega_{i}\right)\right|^{2} d\omega_{s}d\omega_{i}\right)\times N_{p}\label{eq:pairgenrate}
\end{equation}

where $N_{p}=$number of pump photons entering the crystal per second.

\subsection{Time stamping of single and background photons}

\begin{figure}[H]
\includegraphics[width=\linewidth]{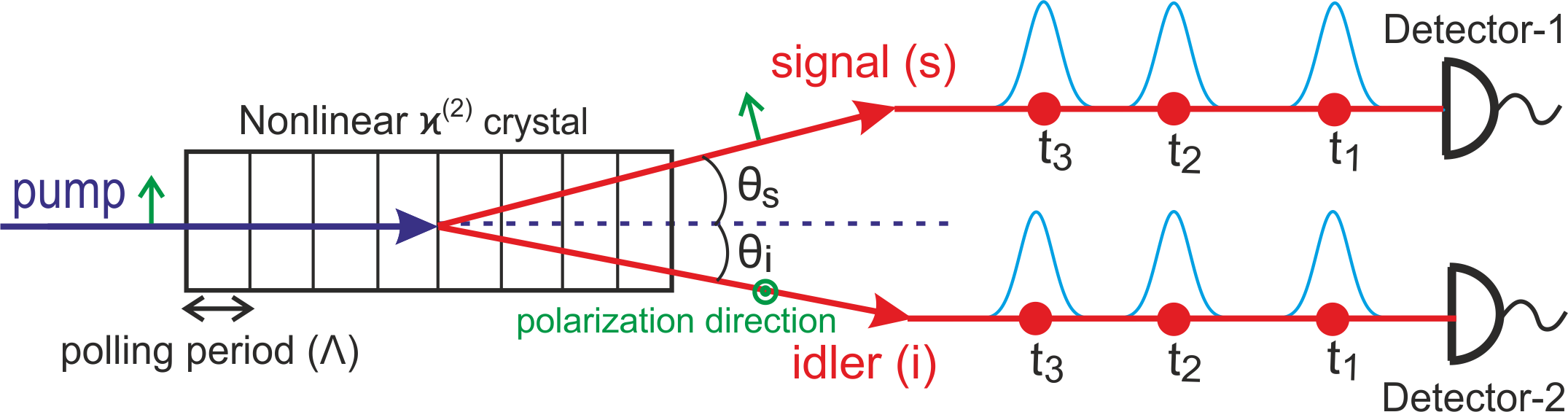}
\caption{A probabilistic ideal single photon source emitting a stream of photon pairs at times $t_{1}, t_{2}, t_{3}$. Detection of the idler photon heralds the signal photon of the same pair. The detection time uncertainty for each photon has been depicted with a Gaussian distribution.}
\label{fig:sketch1}
\end{figure}

In an idealistic picture, we consider a photon source that generates perfect (only) single photon events $\left( \text{Fock state} \left|1\right\rangle\right)$. The quantum uncertainty of detecting each output photon with frequency $\omega$ at time $t$ is $\Delta t \Delta\omega \geq \frac{1}{2}$. The probability distribution of measuring a given photon at time $t$ (or with frequency $\omega$) then becomes the modulus square of its wave function (or its probability amplitude). The distribution also depends on the source properties as well as the filtering conditions. For example, in a SPDC source (as presented through a schematic in Figure\,\ref{fig:sketch1}), the probability distribution can be Gaussian or Sinc-squared, depending on the non-linearity profile of the crystal as well as the spectral profile of the filter. \\

\begin{figure}[!ht]
\includegraphics[width=\linewidth]{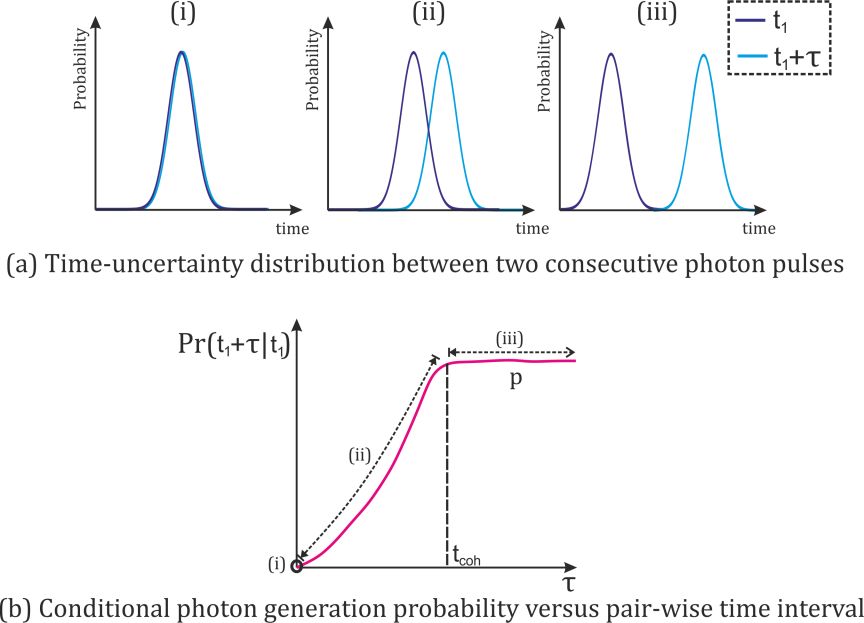}
\caption{(a) Schematics comparing the positions of time-uncertainty distributions for two single photon events with three different relative time intervals $\tau$: (i) $\tau=0$ where $\text{Pr}\left(t_{1}+\tau|t_{1}\right)=0$ (ii) $ 0 < \tau < t_{coh}$ where $0<\text{Pr}\left(t_{1}+\tau|t_{1}\right)<p$ (iii) $\tau\gg t_{coh}$ where $\text{Pr}\left(t_{1}+\tau|t_{1}\right)=p$. (b) A schematic of the nature of the probability distribution $\text{Pr}\left(t_{1}+\tau|t_{1}\right)$ as a function of $\tau$ for any arbitrary $t_1$.}
\label{fig:phstat1}
\end{figure}

Defining the probability of a photon generated at time $t_{1}+\tau$ is $\text{Pr}\left(t_{1}+\tau|t_{1}\right)$ given that the earlier photon was generated at time $t_{1}$; then for an ideal single photon source, for $\tau=0$, this probability becomes zero i.e. $\text{Pr}\left(t_{1}|t_{1}\right)=0$. Now for all $\tau\gg t_{coh}$, where $t_{coh}$ is our coherence time, this probability will have a constant value $p$ i.e. $\text{Pr}\left(t_{1}+\tau|t_{1}\right) = p$. This is because of the fact that for very large values of $\tau$ beyond the coherence time $t_{coh}$, the source behaves truly randomly and emits single photons at any arbitrary interval with equal probability. Therefore as shown in Figure\,\ref{fig:phstat1}(b), if we plot this probability as a function of $\tau$, it will smoothly increase from $0$ and saturate at $p$.\\

Let's assume that $10^{7}$ photons are generated per second. So, we divide the time span of 1 second into equal bins of 1 ps time resolution and then we get $10^{12}$ bins in 1 second. When $10^{7}$ photons are randomly distributed into those $10^{12}$ bins, the probability of having a photon in each bin becomes $10^{-5}$. However, we enforce a restriction that if one photon has already been assigned to a bin, say $x$, then the probability of another photon to be assigned to the same bin ($x$) is zero. Now as illustrated earlier in Figure\,\ref{fig:phstat1}, this probability of assignment for each bin slowly increases for the subsequent bins i.e. $x+1, x+2, x+3,...$ and saturates at $p=10^{-5}$ for some bin number, say in this case $x+n$, where $n$ is related to the coherence time. Please refer to Appendix C for the 
comparison between the simulated and the experimental time stamping data.\par
While generating the time stamps list for each run of the protocol, it has been observed that the process increases the run-time of the simulation significantly. In order to speed up the process, we adapted data re-sampling technique to generate time stamps list. For an arbitrary simulated single photon pair generation rate, time stamps list is generated only for the first instance of the simulation. In further runs of the simulation, where the parameters for the type-II SPDC source module remain unchanged and hence similar pair rate is obtained, the re-sampling technique is used to generate the time stamps list. The application of the re-sampling technique results in variation in the number of time stamps generated for a fixed time-interval. The number of time stamps generated determines the effective number of photon pairs that forms the input to the subsequent modules in the simulation toolkit. In Section\, \ref{subsec:assumptions}, we mentioned that it has been assumed that all the parameters remain invariant over the run-time of the simulation. Thus, for multiple runs of the simulation with fixed parameters, the simulated pair generation rate remains constant over all such instances. But the effective number of photon pairs generated  varies because of the usage of re-sampling technique.


\subsection{Detection of background photons}


Besides the single photons from a SPDC source, the experimental detections also consists of other background photons that commonly originate from any surrounding thermal source. From literature, the photon number (n) distribution for such a source is commonly \emph{super-Poissonian} where $\text{Pr}\left(n\right)=\frac{\mu^n}{\left(1+\mu\right)^{n+1}}$. If we measure the second-order coherence $g^{\left(2\right)}$ for such a source, then it exhibits bunching property, where $g^{\left(2\right)}$ at $\tau=0$ is greater than that at $\tau \gg 0$. 
In our simplified model to simulate these background photon statistics we have considered multi-photon events only up to two photons. Also, we have assumed that $\text{Pr}\left(n=2\right)= \text{Pr}\left(n=1\right)^2$, which implies that $\frac{\text{Pr}\left(2\right)}{\text{Pr}\left(1\right)} \ll 1$ since $0\leq \text{Pr}\left(n\right)\leq 1$. Here, it is important to note that for an ideal single photon distribution $\frac{\text{Pr}\left(2\right)}{\text{Pr}\left(1\right)} \approx 0$.


We assume that $10^{5}$ photons are generated per second. As earlier, we divide the time span of 1 second into equal bins, each of 1 ps time resolution, and then we get $10^{12}$ bins in 1 second. 
Now, let us consider, the probability of assigning a single photon event to an empty bin is $P_1$, then the probability of assigning a multi-photon event to an empty bin becomes $P_1^2$ from the above idea. Therefore the probability of assigning atleast one photon in each bin is $P_1\,+\,2 \times P_1^2$. Thus, from the above assumptions that $10^{5}$ photons are generated per second, the value of $P_1$ becomes $\approx 10^{-7}$. 
We use the above considerations to generate the time stamping data for our background contributions from various thermal sources.\par

Initially, an interpolation data set is generated by varying the incident background rate at the detectors and obtaining the corresponding background coincidence rate for a fixed signal rate. In our case, we have fixed the signal rate at 18 MHz which corresponds to the single photon pair generation obtained from the source simulation for a PPKTP crystal of length 20 mm. The interpolation data set is then pre-stored and used to obtain the incident background rate for any arbitrary input background coincidence rate.\par
The data set for interpolation of the incident background rate from a given background coincidence rate is generated for a specific signal rate of 18 MHz as mentioned. For an arbitrary signal rate, we want to estimate the ratio in which the interpolated background rate will be affected. For a fixed incident background rate, this ratio can be calculated by dividing the background coincidence rate at any arbitrary input signal rate divided by the the background coincidence rate at the 18 MHz signal rate. The final incident background rate is obtained by performing interpolation on the pre-stored interpolation data-set that has been generated, at a fixed background incidence rate, by varying the signal rate and obtaining the background noise level in the coincidence plots for the diagonal and rectilinear basis measurements.




\subsection{Propagation of single photons}

In the source module, each of the generated photons are associated with a Gaussian distribution corresponding to their electric field amplitude. The propagation of single photons have thus been simulated by using the Huygen's principle for propagation of Gaussian beams in 2-dimensions.  

\subsection{Fibre Coupling}

In the experimental set up, coupling of single photons into a single mode fibre is done using stages and couplers that provide certain degrees of freedom to align the fibre-tip with the input beam in order to maximise the coupling. In our simulation toolkit, the fibre coupling sub-module is used to simulate active coupling mechanism that a user might perform while aligning the experimental set up. As mentioned in Section\, \ref{subsec:assumptions}, the position of the fibre tip is simulated accurately for the maximum coupling up to the least count of the coupling apparatus. The error in the position of the fibre tip is simulated by randomly selecting a value from a Gaussian distribution with the least count being twice the sigma of the distribution. The sub-module takes in as input the specifications of the physical components such as the  fibres and the aspheric lenses within the couplers and  the electric field distribution of the incident photons and returns the effective coupling efficiency as the output. The efficiency is calculated as the overlap between the intensity distribution of the received photons at the fibre-tip and the Gaussian distribution with the mode field radius of the fibre as the standard deviation. By further considering the associated losses, the effective coupling efficiency is obtained.

\section{Modelling physical components}
\label{sec:components}
In this section, we will discuss briefly the various physical components that have been simulated as listed in Table\,\ref{tab:components}. The physical components that forms the experimental setup for demonstration of the B92 protocol includes optical components such as lenses, filters and beam splitters, free-space and optical fibre based channel, detectors and TCSPCM. The simulation methodology for the sub-modules corresponding to these physical components are discussed as follows.

\subsection{Lens} 
The working of a lens is simulated by using the lens transfer function calculated from the specifications of the respective lens and applying that on the electric field corresponding to the incident photons. As specified in the list of assumptions, the photons are considered to have the distribution of intensity as Gaussian, so the effect of the lens is simulated in the case of Gaussian beams only. The lens sub-module takes in as input the specifications corresponding to the type of material, the radius of curvatures, etc. and the electric field distribution of the incident photons, and returns the electric field distribution of the photon after it passes through the lens as output.
\subsection{Half-wave plate (HWP)}
The HWP sub-module simulates the the effect of phase shift on the polarization of the photons transmitted through the HWP. The orientation of the fast-axis of the HWP with the respect to the polarization of the incoming photons is simulated by the choice of basis for the required projection. The accuracy of the orientation is limited by the least count of the mounting component and the error is simulated by randomly choosing a value from a Gaussian distribution with the least count being twice the standard deviation of the distribution. The parameters regarding the loss through the medium, least of count of the mounting component etc. are set within the sub-module.
\subsection{Filter} 
The filter sub-module simulates the effect of filter on a beam or photons of certain wavelength. For now we have modelled a very simplified filter inspired from the real components used in the experimental setup and thus can be further enhanced to capture more practical scenarios. The sub-module takes into account the insertion and the transmission losses incurred by the photons incident on the filter.
\subsection{Polarizing beam splitter (PBS)}
The PBS sub-module simulates the transmission of a linearly polarised beam through a PBS. The extinction ratio and the loss associated with the PBS is set within the sub-module as set parameters by taking data from the specifications sheet provided by the manufacturer.
The sub-module takes as input the time stamps of the incident photons and the polarization angle of the photons and returns the time stamps of the photons exiting the transmission and the reflection arm of the PBS in separate lists.

\subsection{Beam splitter (BS)}
The BS sub-module simulates the transmission of a  beam through a beam splitter(PBS). The loss associated with the BS is set within the sub-module as set parameters by taking data from the specifications sheet provided by the manufacturer. The sub-module takes as input the time stamps of the incident photons and returns the time stamps of the photons exiting the transmission and the reflection arm of the BS in separate lists.

\subsection{Fibre transmission}
The fibre-transmission sub-module simulates the transmission of a string of photons through a fibre of a certain length. The sub-module takes into account only the losses encountered by the beam while transmission and does not consider any other effect due to the fibre. This takes as input the length of the fibre channel in metres and the time stamps of the coupled photons into the fibre. The loss associated with the transmission as well as the insertion loss at the mating sleeves are taken from the specification sheet of the fibre and defined within the sub-module. The sub-module finally returns the time stamps of the photons that have reached the exit port of the channel.

\subsection{In-lab free-space transmission}
This sub-module simulates the transmission of a string of photons through free-space of a certain length. It takes into account only the losses encountered by the beam while transmission and does not consider any other effect due to the free-space channel. The sub-module takes in as input the length of the free-space channel in metres and the time stamps of the photons travelling through the channel and returns the time stamps of the photons that have reached the exit port of the channel.

\subsection{Fibre-based single photon detectors}
 This sub-module simulates the detection process of single photons with fibre-based detectors. In other words, we consider the case where single photons are coupled to the detector with a fibre and get detected via avalanche breakdown process at the photo diode of the detector. For each detection event, the detector outputs a TTL pulse corresponding to the detected photon(s). The sub-modules provides a simplistic approach towards modelling of single photon avalanche detectors (SPADs) and hence, is restricted to only SPADs. This takes the time stamps of the stream of photons incident at the detector as an input. The primary detector imperfections that we have considered for simulation are the quantum efficiency of the detector, dead time and timing jitter of the detector, are explained as follows. The values corresponding to these parameters are taken from the specification sheet of the detector and set within the sub-module.
\begin{enumerate}
\item\textit{Quantum efficiency}: A single photon detector has a certain efficiency of detecting photons incident on it i.e. for each of the photons received at the detector, a TTL pulse is not generated. The probability of generation of the TTL pulse on receiving an incident photon is quantified using the parameter quantum efficiency of the detector. The quantum efficiency of the detector or the detector efficiency depends on the wavelength of the incident photons.
\item\textit{Detector dead time}: The dead time of a detector is the time interval after a detection event, followed by an avalanche breakdown, during which the detector is unresponsive to any photon incident at the detector. It defines the time required by the detector to restore the quenching circuit. Thus the minimum time interval possible between two detection events is the dead time of the detector.
\item\textit{Timing jitter}: Due to imperfections in the detector circuit, there is a time  uncertainty between receiving a photon at the detector and generating the TTL pulse. The time interval between generation of the TTL pulse corresponding to a photon detection and the time at which the photon is actually received at the detector is not constant and has a Gaussian distribution. This uncertainty is quantified by the timing jitter of the detector, i.e. the full width at half maximum (FWHM) of the distribution.
\end{enumerate}
\par
The logic used for simulating the single photon detection event at the detectors considering the specified imperfections can be explained as follows.
The quantum efficiency affects the difference in the  total number of photons incident at the detector and those actually detected. For each photon incident at the detector, we generate a random number in the range $\left[0,1\right]$ as in principal the efficiency of the detector is $\leq 1$. If the random variable has a value less than the quantum efficiency of the detector, the instance is considered as the detection event.
The detector dead time is then accounted by checking the difference in the time stamp value of the consecutive photons received at the detector. If the difference is less than the dead time, the photon corresponding to the larger time stamp value is discarded and the time difference with the next photon is checked and the process continues for all the received photons. To simulate the timing jitter of the detector, each of the generated TTL pulses corresponding to detected photons is adjusted with a time delay chosen randomly from a Gaussian distribution with a mean ($\mu$) of zero and standard deviation ($\sigma$) value equalling to 2.3 times the timing jitter of the detector.

\subsection{Time Correlated Single Photon Counting Module (TCSPCM)}
A TCSPCM receives the TTL pulses generated at the detector corresponding to the detection events and registers time stamps for the TTL pulses. The TCSPCM sub-module simulates this process by taking as input the time stamps of the TTL pulses generated at the detector corresponding to the detection events. The primary TCSPCM imperfections that we have considered for simulation are the losses in the SMA cables that connects it to the detectors, dead time and timing jitter of the TCSPCM.
In the following, we explain the various parameters that have been accounted to model the TCSPCM.
\begin{enumerate}
\item \textit{SMA cable losses}:  SMA cables are used to connect detector to the TCSPCM. These cables have some inherent losses at the connectors and as a result of that, some of the TTL pulses generated from the detectors are lost.
\item \textit{TCSPCM dead time}: Similar to the detectors, the dead time of a TCSPCM is the  minimum time interval between registering two TTL pulses received from the detector. The value of the dead time is set within the sub-module as per the specification sheet.
\item \textit{Timing jitter}: Due to imperfections in the TCSPCM circuit,  the time stamps associated to a received pulse is not the same as the time at which the TTL pulse was received. The distribution of this time delay is a Gaussian distribution, centred at the time at which the pulse was actually received by the TCSPCM. This deviation is quantified using the timing jitter of the TCSPCM.
\end{enumerate}

The dead time of the TCSPC module and the timing jitter is simulated in the same way as that has been done for the detector. For the SMA cable losses, the channel efficiency is calculated using the relation: $tcspcm_e=10^{-\frac{loss}{10}}$.

\section{Results and discussion}
\label{sec:results}

Our manuscript introduces a new simulation toolkit called qkdSim. While in the future, we aim to develop this into a software that will be able to simulate any QKD protocol along with consideration of the associated experimental imperfections, in the present work, we show details of the B92 protocol simulation and the comparison of its performance analysis with the results of our experimental demonstration of B92 protocol using a heralded single photon source. We find a reasonably good match between theory and experiment as is discussed below.\par


Here we compare the final results obtained from the experiment with those from simulations and then discuss their implications. The comparative results from the experiment and simulations, using the two optimization strategies A and B, are presented in Tables\,\ref{tab:resultTableA} and \ref{tab:resultTableB} respectively. The results reported for the experiment have been obtained by averaging the measured outcomes over 20 measurement sets, wherein each set involved a measurement time of 10 seconds. For the simulation, while number of sets for averaging remained the same, the runtime of the protocol was considered to be 1 second. Here, the consideration of a smaller runtime for the simulation can be motivated from the assumption of time-invariance that has been considered for this system. The error values quoted are the standard deviations of the 20 iterations respectively. 
\begin{center}
\begin{table}[H]
\centering
\begin{tabular}{|c|c|c|c|c|c|c|c|}
\hline
\multirow{3}{*}{\begin{tabular}[c]{@{}c@{}}Crystal\\ length\\ (mm)\end{tabular}} & \multirow{3}{*}{\begin{tabular}[c]{@{}c@{}}Time\\ of\\ the\\ day\end{tabular}} & \multicolumn{6}{c|}{Optimization strategy A}                                                                                                            \\ \cline{3-8} 
                                                                                 &                                                                            & \multicolumn{3}{c|}{From experiment}                                                                                                                                      & \multicolumn{3}{c|}{From simulation}                                                                                                                                      \\ \cline{3-8} 
                                                                                 &                                                                            & \begin{tabular}[c]{@{}c@{}}key\\ rate\\ (kHz)\end{tabular} & \begin{tabular}[c]{@{}c@{}}QBER\\ (\%)\end{tabular}  & \begin{tabular}[c]{@{}c@{}}\makecell{asym-\\metry\\ (\%)}\end{tabular} & \begin{tabular}[c]{@{}c@{}}key\\ rate\\ (kHz)\end{tabular} & \begin{tabular}[c]{@{}c@{}}QBER\\ (\%)\end{tabular}  & \begin{tabular}[c]{@{}c@{}}\makecell{asym-\\metry\\ (\%)}\end{tabular} \\ \hline
\multirow{2}{*}{20}                                                              & Day                                                                        & \begin{tabular}[c]{@{}c@{}}47.6\\ $\pm$0.6\end{tabular}     & \begin{tabular}[c]{@{}c@{}}4.79\\ $\pm$0.01\end{tabular} & \begin{tabular}[c]{@{}c@{}}49.82\\ $\pm$0.01\end{tabular}    & \begin{tabular}[c]{@{}c@{}}53.08\\ $\pm$0.31\end{tabular}        & \begin{tabular}[c]{@{}c@{}}4.79\\ $\pm$0.01\end{tabular} & \begin{tabular}[c]{@{}c@{}}50.1\\ $\pm$0.06\end{tabular}      \\ \cline{2-8} 
                                                                                 & Night                                                                      & \begin{tabular}[c]{@{}c@{}}51.0\\ $\pm$0.5\end{tabular}     & \begin{tabular}[c]{@{}c@{}}4.79\\ $\pm$0.01\end{tabular} & \begin{tabular}[c]{@{}c@{}}50.15\\ $\pm$0.02\end{tabular}    & \begin{tabular}[c]{@{}c@{}}52.83\\ $\pm$0.36\end{tabular}        & \begin{tabular}[c]{@{}c@{}}4.79\\ $\pm$0.01\end{tabular} & \begin{tabular}[c]{@{}c@{}}50.1\\ $\pm$0.05\end{tabular}      \\ \hline
\multirow{2}{*}{30}                                                              & Day                                                                        & \begin{tabular}[c]{@{}c@{}}33\\ $\pm$2\end{tabular}         & \begin{tabular}[c]{@{}c@{}}4.78\\ $\pm$0.01\end{tabular} & \begin{tabular}[c]{@{}c@{}}50.07\\ $\pm$0.02\end{tabular}    & \begin{tabular}[c]{@{}c@{}}64.11\\ $\pm$0.98\end{tabular}        & \begin{tabular}[c]{@{}c@{}}4.78\\ $\pm$0.01\end{tabular} & \begin{tabular}[c]{@{}c@{}}50.05\\ $\pm$0.08\end{tabular}      \\ \cline{2-8} 
                                                                                 & Night                                                                      & \begin{tabular}[c]{@{}c@{}}36\\ $\pm$3\end{tabular}         & \begin{tabular}[c]{@{}c@{}}4.78\\ $\pm$0.01\end{tabular} & \begin{tabular}[c]{@{}c@{}}50.08\\ $\pm$0.02\end{tabular}    & \begin{tabular}[c]{@{}c@{}}59.56\\ $\pm$1.55\end{tabular}        & \begin{tabular}[c]{@{}c@{}}4.79\\ $\pm$0.01\end{tabular} & \begin{tabular}[c]{@{}c@{}}50.01\\ $\pm$0.11\end{tabular}      \\ \hline
\end{tabular}
\caption{Optimized results of average key rate, QBER and asymmetry (i.e. key symmetry), obtained using strategy A, from the experiment and the simulation.}
\label{tab:resultTableA}
\end{table}
\end{center}

\begin{center}
\begin{table}[H]
\centering
\begin{tabular}{|c|c|c|c|c|c|c|c|}
\hline
\multirow{3}{*}{\begin{tabular}[c]{@{}c@{}}Crystal\\ length\\ (mm)\end{tabular}} & \multirow{3}{*}{\begin{tabular}[c]{@{}c@{}}Time\\ of\\ the\\ day\end{tabular}} & \multicolumn{6}{c|}{Optimization strategy B}                                                                                                                                                                                                                                                                                                              \\ \cline{3-8} 
&  & \multicolumn{3}{c|}{From experiment}                                                                                        & \multicolumn{3}{c|}{From simulation}                                                                                                                                      \\ \cline{3-8} 
&  & \begin{tabular}[c]{@{}c@{}}key\\ rate\\ (kHz)\end{tabular} & \begin{tabular}[c]{@{}c@{}}QBER\\ (\%)\end{tabular}  & \begin{tabular}[c]{@{}c@{}}\makecell{asym-\\metry\\ (\%)}\end{tabular} & \begin{tabular}[c]{@{}c@{}}key\\ rate\\ (kHz)\end{tabular} & \begin{tabular}[c]{@{}c@{}}QBER\\ (\%)\end{tabular}  & \begin{tabular}[c]{@{}c@{}}\makecell{asym-\\metry\\ (\%)}\end{tabular} \\ \hline
\multirow{2}{*}{20}                                                              & Day                                                                        & \begin{tabular}[c]{@{}c@{}}47.8\\ $\pm$0.6\end{tabular}     & \begin{tabular}[c]{@{}c@{}}4.79\\ $\pm$0.01\end{tabular} & \begin{tabular}[c]{@{}c@{}}50.2\\ $\pm$0.3\end{tabular}      & \begin{tabular}[c]{@{}c@{}}59.97\\ $\pm$0.25\end{tabular}        & \begin{tabular}[c]{@{}c@{}}4.79\\ $\pm$0.01\end{tabular} & \begin{tabular}[c]{@{}c@{}}56.95\\ $\pm$0.2\end{tabular}          \\ \cline{2-8} 
                                                                                 & Night                                                                      & \begin{tabular}[c]{@{}c@{}}53.8\\ $\pm$0.4\end{tabular}     & \begin{tabular}[c]{@{}c@{}}4.79\\ $\pm$0.01\end{tabular} & \begin{tabular}[c]{@{}c@{}}53.7\\ $\pm$0.3\end{tabular}      & \begin{tabular}[c]{@{}c@{}}59.81\\ $\pm$0.25\end{tabular}        & \begin{tabular}[c]{@{}c@{}}4.79\\ $\pm$0.01\end{tabular} & \begin{tabular}[c]{@{}c@{}}56.97\\ $\pm$0.19\end{tabular}          \\ \hline
\multirow{2}{*}{30}                                                              & Day                                                                        & \begin{tabular}[c]{@{}c@{}}36\\ $\pm$2\end{tabular}         & \begin{tabular}[c]{@{}c@{}}4.79\\ $\pm$0.01\end{tabular} & \begin{tabular}[c]{@{}c@{}}54.0\\ $\pm$0.3\end{tabular}      & \begin{tabular}[c]{@{}c@{}}71.42\\ $\pm$0.95\end{tabular}        & \begin{tabular}[c]{@{}c@{}}4.79\\ $\pm$0.01\end{tabular} & \begin{tabular}[c]{@{}c@{}}57.05\\ $\pm$0.32\end{tabular}          \\ \cline{2-8} 
                                                                                 & Night                                                                      & \begin{tabular}[c]{@{}c@{}}38\\ $\pm$3\end{tabular}         & \begin{tabular}[c]{@{}c@{}}4.78\\ $\pm$0.01\end{tabular} & \begin{tabular}[c]{@{}c@{}}54.1\\ $\pm$0.4\end{tabular}      & \begin{tabular}[c]{@{}c@{}}65.86\\ $\pm$1.8\end{tabular}        & \begin{tabular}[c]{@{}c@{}}4.79\\ $\pm$0.01\end{tabular} & \begin{tabular}[c]{@{}c@{}}57.11\\ $\pm$0.3\end{tabular}          \\ \hline
\end{tabular}
\caption{Optimized results of average key rate, QBER and asymmetry (i.e. key symmetry), obtained using strategy B, from the experiment and the simulation.}
\label{tab:resultTableB}
\end{table}
\end{center}

All the reported results have been obtained using the optimization strategies (introduced in Section\,\ref{subsec:dataanalysis}) and as a consequence of that the results are dependent on the background noise in the coincidence plots. The source of this noise comes from the background photons incident at the detector. Firstly, we observe in Table\,\ref{tab:resultTableA} that the error values for both the estimated QBER and asymmetry parameters, are at least one order of magnitude lower than those for the key rates, irrespective of their origin: i.e. from experiment or simulation. The reason for this being the constraints on QBER ($\approx 4.8\%$) and key symmetry ($\approx 50\%$), which were fixed in course of the optimization strategy A to infer the estimated key rate, as discussed earlier in Section\,\ref{subsec:dataanalysis}. Therefore, all the fluctuations in the measured data get reflected on the estimated key rate figures. However, in Table\,\ref{tab:resultTableB}, we observe that both the key rate and key symmetry (or asymmetry in the key) have one order of magnitude higher error values than QBER, as in optimization strategy B asymmetry was also considered an unconstrained parameter for optimization along with key rate. It is important to note that such choice of optimization results involves different implications on the security of the resultant secure key as discussed earlier in Section\,\ref{subsec:dataanalysis}.

Secondly, for both tables, our in-lab free-space experimental demonstration of the B92 protocol (discussed in Section~\ref{subsec:exp}) was conducted during day and night time. For both time periods, the similarity of the experimental results reflects that the in-lab conditions varied indistinctly at different times of the day. In principle, the estimated key rate should be same for both the day and the night time measurements; however, we hypothesize that the thermostatic regulations and other stray light shielding facilities, having realistic imperfections, introduce the increase in key rate estimates for the night time measurements. As the background noise level forms a key input to the simulation toolkit, the negligible variation of the in-lab conditions at different times of the day gets reflected in the simulation results as well. \par

Thirdly, again for both tables in general, we note that though having a longer crystal length should potentially lead to a higher key rate due to increased pair production rate, the 30 mm crystal length produces a lower key rate estimate than the 20 mm one  for the experimental results unlike the simulation part, where the logic is rather consistent. This is because in the experimentally measured data sets, the 30 mm crystal besides having higher signal level also posses an increased noise level due to the detection of more background photons, which then lowers the key rate to ensure that the QBER optimization remains within the threshold value of 4.8\%. However, by the same logic, the simulation should also have a lower key rate than that for the 20 mm crystal length, since the same background rate forms an input to the simulation. Nevertheless, we observe a contradiction there. As per our understanding, the origin of this discrepancy is due to the assumptions involved with the pair generation rate calculation while simulating the type-II collinear degenerate SPDC process. The simulated pair generation rate is directly proportional to the crystal length as well as the input pump intensity at the crystal which wasn't observed experimentally. Through separate experimental tests we have verified that the singles and the coincidence rates observed at the detectors does not increase linearly with increase in the pump beam intensity and the crystal length. From these test results, it can be inferred that the pair generation rate at the crystal does not increase linearly as well. Thus the pair generation rate obtained from the simulation differs from the actual value obtained from the experiment, resulting in discrepancy between the final results for the 30mm crystal length. Additionally, in the case of the experimentally estimated key rate with the 30 mm crystal, the error values are also higher owing to the increased fluctuations in the measured data points on the coincidence plot. In a nutshell, due to these reasons, the results from the experiment and those from the simulation offer a better match when the length of the crystal is 20 mm compared to the 30 mm case that involves both higher background noise and more fluctuations. \par


Lastly, it is important to point out that while the optimization strategies focus on fixed values of QBER and asymmetry in the key-string, the estimated key rate is itself associated with a standard deviation in case of both the experiment and simulation. While the sources of this deviation for the experiment are imperfections of the source, devices, components etc., that for the simulation are primarily captured by the methodology of the simulation which is based on random number generation from both uniform and normal distributions. While certain aspects of the experiment such as loss through the medium, generation of time stamps and detection efficiency are simulated using random number generation from a uniform distribution; the alignment errors, timing jitter, etc. have been simulated using random numbers from a normal distribution. This, explained in detail in Sections\,\ref{sec:process} and \ref{sec:components}, results in deviation in the outcomes for multiple simulation runs of the protocol which have been shown in the tables.

\section{Conclusion and outlook}
\label{sec:conclusion}

In this manuscript, we discuss in detail our in-house developed simulator qkdSim, which has been created to specifically provide the QKD community with a simulation toolkit that takes into account practical imperfections as may be encountered in an actual experiment. Earlier available softwares including qkdX do not contain much discussion on attendant physical processes and/or physical components. Our qkdSim aims to bring in more practical considerations to QKD simulations so that realistic predictions about the key rate and the QBER can be made before investment of resources in developing the physical QKD system. To this end, in our current work, we have shown how qkdSim simulates the B92 protocol in detail. We have discussed simulations of various physical components as well as physical processes following the Agifall software model. A representative key rate from the experiment is $51 \pm 0.5$ Kbits/sec whereas that from the qkdSim simulated value is $52.83 \pm 0.36$ Kbits/sec, corresponding to a representative QBER of $4.79\% \pm 0.01\%$ from both. Having successfully simulated the B92 protocol, which is an example of a QKD protocol that does not use entanglement as the basis for security, we will, in future work address the applicability of qkdSim to entanglement based QKD. This will bring us a step closer to the desired all-purpose QKD software, which is capable of simulating arbitrary QKD protocols, giving due importance to experimental imperfections and conditions.\\

\section*{ACKNOWLEDGEMENTS}

US would like to thank the Indian Space Research Organisation for support through the QuEST-ISRO research grant.

\bibliography{output}


\appendix

\begin{widetext}

\section*{Appendix A: Bootstrapping on measurement outcomes}
\label{app:bootstrap}

\subsection*{Motivation}

To estimate an optimal choice for the data acquisition rate (measurement time for each data set versus number of data sets measured) for an appropriate estimation of key rate, QBER and asymmetry (or key symmetry) obtained in our experimental demonstration of the B92 protocol.

\subsection*{Methodology}

For a given measurement run of \textsc{totDatasets} sets, each of time \textsc{totTime} picoseconds:

\begin{enumerate}

\item We choose the start and end time window marker positions for both the coincidence plots i.e. between Alice \& Bob \textbf{V} basis (vertical) and Alice \& Bob \textbf{+} basis (diagonal). This provides us with three lists of time stamps sorted in ascending order corresponding to each measurement set.

\item We choose an uniform random sequence of length \textsc{totDatasets}.

\item Using a fast binary search algorithm we collect $k$ (<\textsc{totTime}) seconds of data from each of the above three lists, starting from the time stamp entry found in the random sequence \textsc{sequence}. For those three output lists we calculate the Keyrate.

\item We repeat step-3 for each element of \textsc{sequence} and then obtain the average Keyrate over those elements.

\item To remove any bias, we randomize (or repeat) steps 2-4 over many iterations. We store the average Keyrate, QBER and Asymmetry obtained in each iteration.

\item We finally calculate the ratio: standard deviation (SD) over mean (M) of all the average Keyrates.

\item We repeat steps 2-6 for a range of $k$ values and finally obtain a plot over that range.
\end{enumerate}

\subsection*{Result}

We tested our bootstrap method 
over a measurement run where (\textsc{totDatasets}=) 20 sets of (\textsc{totTime}=) $15\times 10^{12}$ picoseconds data were collected. In order to remove bias, for each set and corresponding to every chosen size of time window, the estimated values of key rate, QBER \& key symmetry were averaged over 10000 iterations. The result obtained is presented in Figure\,\ref{fig:bootstarp_result-3}.
\begin{center}
\begin{figure}[H]
\centering
\includegraphics[scale=0.4]{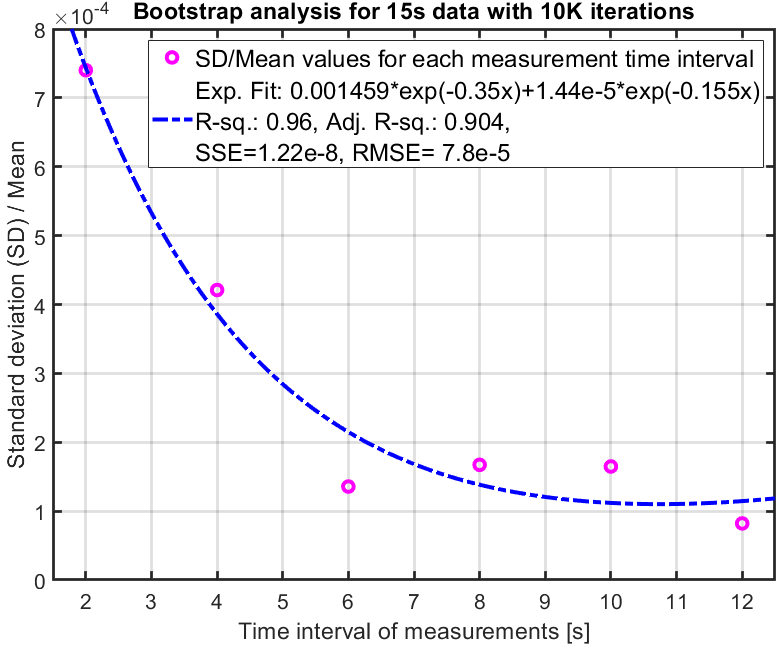}
\caption{Bootstrap analysis (SD/Mean of key rate) plot on a 15 seconds data-set from a series of 20 measurement runs. Here each run consists of 10000 iterations. The saturation or convergence effect is supported by the exponential fit to the simulated data-points. The goodness of the fit is gauranteed by the low values for sum-of-squares error (SSE) and root-mean-square error (RMSE) as well as close to 1 values for R-squared (R-sq.) and adjusted R-squared (Adj. R-sq.).}
\label{fig:bootstarp_result-3}
\end{figure}
\par\end{center}
As expected, from this result, we observe that the ratio of SD by Mean decreases with increase in the data collection time of $k$ seconds for each measurement and finally saturates beyond a certain value (here say, $\approx 9$ seconds) of $k$ as it approaches \textsc{totTime}. To have an appropriate estimate of the key rate, QBER and key symmetry, the measurement runtime (which for our experiment was fixed to 20 sets of 10 seconds each) should belong to this saturation region.

\section*{Appendix B: Optimization methods for data analysis}
\label{app:optimzationmethods}

\subsection*{Overview}

In classical cryptography, \textbf{asymmetry} quantifies the disparity between the number of \enquote{0} bits and \enquote{1} bits in the key shared by Alice and Bob. For perfect secrecy of the key string, the probability of a certain key bit to be \enquote{0} or \enquote{1} should be equal for all the bits in the key string. Thus, we can consider that asymmetry in key string can give rise to security issues in QKD protocols. In the implementation of a QKD protocol, the sifted key generated can be asymmetric since the various optical components induce some imperfections.\par 

To account for this asymmetry in key string due to device imperfection, we define two types of optimization strategies, namely A \& B, to obtain the optimal values for the key rate, QBER and key symmetry. Both strategies have their advantages and short-comings. In strategy A, we maximize the key rate while keeping the QBER below a certain threshold and maintaining approximately 50:50 key symmetry. Here, the asymmetry obtained in the key string is negligible, however the fixing of asymmetry value introduces the possibility for leakage of additional information to the eavesdropper. On the other hand in strategy B, we maximize the key rate with similar constraints on QBER but not on key symmetry. With this technique, the key rate gets increased; however now the security gets compromised to some extent since the probability for obtaining any key string out of $2^N$ options, where $N$ is no. of key bits in each key string, remains no longer $\frac{1}{2}$.

\subsection*{Methodology}

In our experimental version of the B92 protocol, we measure two coincidence curves which includes coincidences between: Alice \& Bob's \textbf{V} basis and Alice \& Bob's \textbf{+} basis, as sketched in Figure\,\ref{fig:schematic}.\\
\newline

\paragraph*{Strategy A:} We use the 
this method to maximize the output key rate while maintaining an asymmetry (also referred to as key symmetry) value of $\thickapprox50\%$ and simultaneously ensuring a  quantum-bit-error-rate (QBER) $\leq4.8\%$ in our protocol. For a given dataset in a measurement run $S$ of datasets, the main steps of this strategy are listed as follows:
\begin{enumerate}
\item Detect and mark the coincidence maximum points in both plots.

\item In the first plot, consider an almost wide window, i.e. place the window markers on the left (or \textsc{winLeftCurve1}) and the right (or \textsc{winRightCurve1}) of the coincidence maximum. Ensure that they are located far beyond the full-width-at-half-maxima (FWHM) points. 

\item Move both the window markers with equal step size towards the coincidence maximum and in turn maximize the key rate (i.e. whole area under the curve) within the considered window span. Ensure that the QBER remains below the threshold value of $4.8$\% during maximization.

\item Retain the optimized window position in the left concidence plot and in a similar way optimize the left (or \textsc{winLeftCurve2}) and the right (or \textsc{winRightCurve2}) window marker positions on the right coincidence plot. During this optimization, ensure that $\thickapprox50\%$ symmetry exists between the key rates obtained from both the curves and also that the overall QBER from both curves lies within $4.8$\%.
\end{enumerate}

Lastly, store the optimized key rates, QBERs and key symmetry (or asymmetry) values for all the $S$ datasets in three different lists: say \textsc{optKey}, \textsc{optQber} and \textsc{optAsymmetry} respectively. Calculate the mean value for each of the three lists to obtain the optimal key rate, QBER and key symmetry for the entire measurement run over $S$ datasets.\\
\newline

\paragraph*{Strategy B:} We use the 
this method to maximize the output key rate while only ensuring a quantum-bit-error-rate (QBER) $\leq4.8\%$ in our protocol. For a given dataset in a measurement run $S$ of datasets, the main steps of this strategy are listed as follows:
\begin{enumerate}
\item Detect and mark the coincidence maximum points in both plots.

\item In the first plot, consider an almost wide window, i.e. place the window markers on the left (or \textsc{winLeftCurve1}) and the right (or \textsc{winRightCurve1}) of the coincidence maximum. Ensure that they are located far beyond the full-width-at-half-maxima (FWHM) points. 

\item Move both the window markers with equal step size towards the coincidence maximum and in turn maximize the signal to noise ratio (SNR), i.e. the ratio of the areas under the curve: above and below the upper bound background noise level, within the considered window span. 

\item Retain the optimized window position in the left concidence plot and in a similar way optimize the left (or \textsc{winLeftCurve2}) and the right (or \textsc{winRightCurve2}) window marker positions on the right coincidence plot. 

\item After SNR optimization on the second coincidence plot, move both its window markers in or out to achieve the current window span on the first coincidence plot.

\item Alter the window size by moving the slightly markers in/out to ensure that the overall QBER doesn't cross the threshold value of $4.8$\%.

\end{enumerate}

In the similar approach as used in strategy A, obtain the mean value from each of the three optimized lists for key rates, QBERs and asymmetry values to report the optimal key rate, QBER and key symmetry for the entire measurement run.

\section*{Appendix C: Single photon time-stamping}
Given the time stamping data for single photons emitted from an ideal single photon source, we plot the distribution for pair-wise time interval, where the X-axis is the time difference between any two consecutive photons ($t_{n+1}-t_{n}$), and Y-axis represents the number of such events per second. This distribution possesses an anti-bunching property at smaller time scales ($\Delta t \rightarrow 0$) (see Figure\,\ref{fig:phresult1}) and an exponential decay nature at larger time scales (far from zero time interval, $\Delta t >> 0$) as depicted in Figure\,\ref{fig:phresult2}.
\begin{figure}[!ht]
\centering
\includegraphics[scale=0.35]{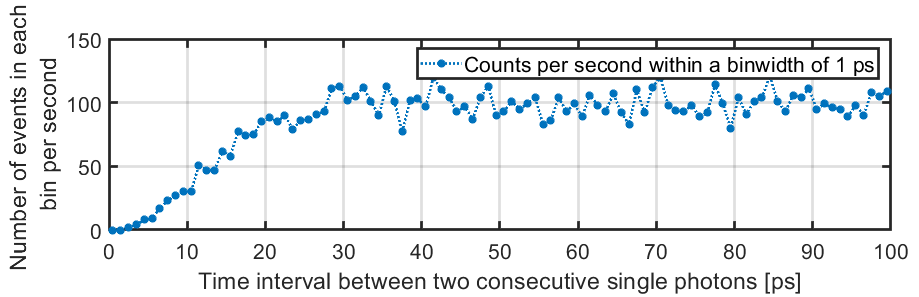}
\caption{\footnotesize Simulation: at high resolution (i.e. narrow bin width of 1 ps), we observe the anti-bunching behaviour of a single photon distribution. The exponential decay-rate being low is invisible at very short (highly resolved) time periods. More particularly, at zero time difference no events occurred since the multi-photon probability was considered to be zero. Here, the standard deviation ($\sigma$) was arbitrarily taken to be 10 ps and so the curve owing to the number of events saturates after $3\sigma$ of pair-wise time difference.}
\label{fig:phresult1}
\end{figure}
\begin{figure}[!ht]
\centering
\includegraphics[scale=0.6]{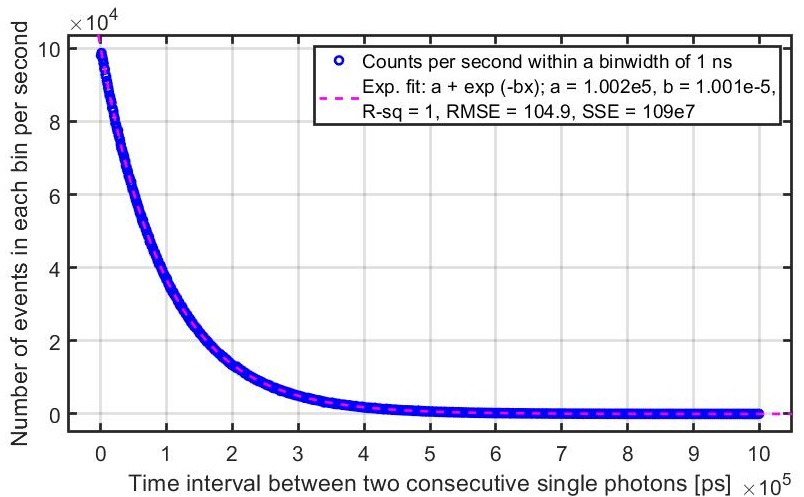}
\caption{\footnotesize Simulation: at low resolution (i.e. broader bin width of 1 ns), we observe the exponential decay nature which is expected from any random distribution. Here, the abbreviations RMSE, SSE and R-sq. stand for root mean square error, sum of square error, and R-squared respectively. They determine the relative and absolute goodness of the fit.}
\label{fig:phresult2}
\end{figure}

This behaviour (exponential decay) was also noticed in our experimental data as shown in Fig~\ref{fig:phexpresult}. We cannot observe the anti-bunching behaviour at shorter time scale as the deadtime (45 ns) of our detector is much larger than the coherence time. 

\begin{figure}[!ht]
\centering
\includegraphics[scale=0.4]{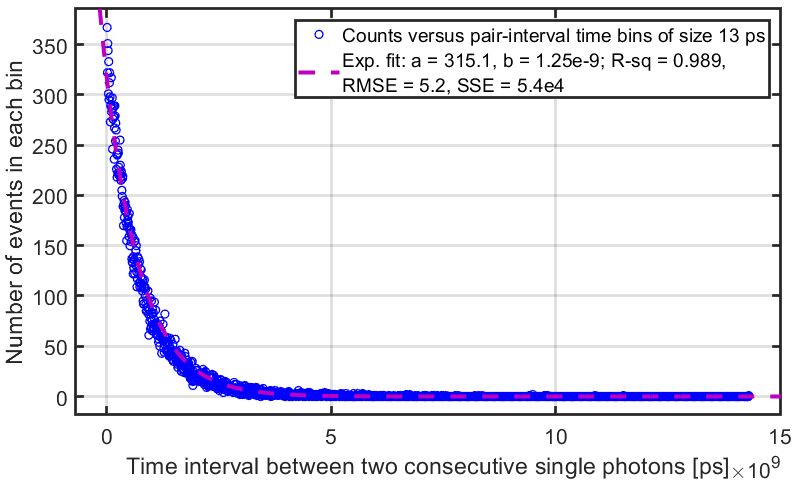}
\caption{\footnotesize Measurement: considering a binsize of 13 ps we observe the exponential decay nature for the distribution of frequency of single photon events versus pair-wise time interval between two consecutive photons emitted from the SPDC source along the signal (or idler) path. The measurements were collected for time window of 2 seconds and a pump power of 2 mW was used. The R-squared (R-sq.) value of the exponential fit in pink is $\approx 1$ ensuring a nice fit to the measured data-points.}
\label{fig:phexpresult}
\end{figure}

\section*{Appendix D: Phase matching temperature for our PPKTP crystal}

Considering the law of momentum conservation (stated in Figure\,\ref{fig:spdc-conservation}) for a SPDC process (described in Figure\,\ref{fig:schematic-spdc}), the phase matching condition for a periodically poled crystal:

\begin{eqnarray}
K_{p}\cos\theta_{p} & = & K_{s}\cos\theta_{s}+K_{i}\cos\theta_{i}+\frac{2\pi}{\Lambda\left(T\right)}\nonumber\\
K_{s}\sin\theta_{s} & = & K_{i}\sin\theta_{i}\label{eq:momconserve}
\end{eqnarray}
 where $\varLambda\left(T\right)$ is the poling period of the crystal
dependent on temperature $T$ and $\theta$ is the angle with respect
to the direction of pump propagation. Further considering the conditions
of colinearity $\left(\textrm{i. e.}\:\theta_{s}=\theta_{i}=0\right)$
and degeneracy $\left(\textrm{i. e.}\;\omega_{s}=\omega_{i}=\frac{\omega_{p}}{2}\right)$,
also $K=\frac{2\pi\,n}{\lambda}$. Thus substituting these conditions
in Eq. \ref{eq:momconserve} we get,

\begin{equation}
\frac{2\pi\,n_{p}}{\lambda_{p}}=\frac{2\pi\,n_{s}}{\lambda_{s}}+\frac{2\pi\,n_{i}}{\lambda_{i}}+\frac{2\pi}{\Lambda\left(T\right)}\label{eq:momconservetwo}
\end{equation}

where, $n_{p}$, $n_{s}$ and $n_{i}$ are the nonlinear refractive indices for the pump, signal and idler photons. Also, $\lambda_{p}$, $\lambda_{s}$ and $\lambda_{i}$ represent the wavelengths of the pump, signal and idler photons respectively.

Now, each of these $n_{h}=f\left(T,\:\hat{s},\:\lambda_{h}\right)$
where $h\,\epsilon\,\left\{ p,\,s,\,i\right\} $; with T being temperature,
$\hat{s}$ being polarization direction and $\lambda_{h}$ being its
wavelength. So, a thermal expansion of the poling period gives \cite{s-emamueli2003},

\begin{eqnarray}
\Lambda\left(T\right) & = & \varLambda_{0}\left\{ 1+\alpha\left(T-T_{0}\right)+\beta\left(T-T_{0}\right)^{2}\right\} \label{eq:polperiodexpand}
\end{eqnarray}

where $T_{0}=25^{\circ}C$ room temperature, also for KTP crystal
$\alpha=\left(6.7\pm0.7\right)\times10^{-6}\left[deg\,C^{-1}\right]$
and $\beta=\left(11\pm2\right)\times10^{-9}\left[deg\,C^{-1}\right]$.
Also, a thermal expansion of refractive indices provides \cite{s-emamueli2003, wiechmann1993},

\begin{eqnarray}
n\left(\lambda,T\right) & = & n\left(\lambda,\:T=T_{0}\right)+\left.\frac{\partial n}{\partial T}\right|_{\left(\lambda,\,T=T_{0}\right)}\left(T-T_{0}\right) \nonumber \\ & + & \left.\frac{\partial^{2}n}{\partial T^{2}} \right|_{\left(\lambda,\,T=T_{0}\right)} \left(T-T_{0}\right)^{2}.\label{eq:riexpand}
\end{eqnarray}

Say, $n_{1}\left(\lambda\right)=\left.\frac{\partial n}{\partial T}\right|_{\left(\lambda,\,T=T_{0}\right)}$
and $n_{2}\left(\lambda\right)=\left.\frac{\partial^{2}n}{\partial T^{2}}\right|_{\left(\lambda,\,T=T_{0}\right)}$. 

Now, Sellmeier equations for PPKTP crystals are \cite{kato2002,bierlein1989,Fradkin1999}:

\begin{eqnarray}
\textrm{one pole: }n^{2}\left(\lambda,T=T_{0}\right) & = & A+\frac{B}{1-C\lambda^{-2}}-D\lambda^{2}; \& \label{eq:onepoleSE}
\end{eqnarray}

\begin{eqnarray}
\textrm{two pole: }n_{z}^{2}\left(\lambda,T=T_{0}\right) & = & A+\frac{B}{1-C\lambda^{-2}}\\\nonumber
& - & \frac{D}{1-E\lambda^{-2}} - F\lambda^{2}. \label{eq:twopoleSE}
\end{eqnarray}

\end{widetext}

\end{document}